\documentclass[article,shortnames,nojss]{jss}

\usepackage{
	amsfonts,
	amsmath,
	booktabs,
	bm,
	todonotes,
    color,
    appendix}

\author{Quentin F. Gronau\\University of Amsterdam \And 
        Akash Raj K. N.\\University of Amsterdam \And 
        Eric-Jan Wagenmakers\\University of Amsterdam}
\title{Informed Bayesian Inference for the A/B Test}

\Plainauthor{Quentin F. Gronau,
Akash Raj K. N.,
Eric-Jan Wagenmakers} 
\Plaintitle{Informed Bayesian Inference for the A/B Test} 
\Shorttitle{\pkg{abtest}} 

\Abstract{Booming in business and a staple analysis in medical trials, the A/B test assesses the effect of an intervention or treatment by comparing its success rate with that of a control condition. Across many practical applications, it is desirable that (1) evidence can be obtained in favor of the null hypothesis that the treatment is ineffective; (2) evidence can be monitored as the data accumulate; (3) expert prior knowledge can be taken into account. Most existing approaches do not fulfill these desiderata. Here we describe a Bayesian A/B procedure based on \citet{KassVaidyanathan1992} that allows one to monitor the evidence for the hypotheses that the treatment has either a positive effect, a negative effect, or, crucially, no effect. Furthermore, this approach enables one to incorporate expert knowledge about the relative prior plausibility of the rival hypotheses and about the expected size of the effect, given that it is non-zero. To facilitate the wider adoption of this Bayesian procedure we developed the \pkg{abtest} package in \proglang{R}. We illustrate the package options and the associated statistical results with a fictitious business example and a real data medical example.}
\Plainkeywords{model comparison, Bayes factor, prior elicitation, Bayesian estimation} 

\Address{
  Quentin F. Gronau\\
  Department of Psychological Methods\\
  University of Amsterdam\\
  Nieuwe Achtergracht 129 B\\
  1018 WT Amsterdam, The Netherlands\\
  E-mail: \email{Quentin.F.Gronau@gmail.com}
}

\begin{document}

\section{Introduction}
Does the modification of a company website increase the number of online purchases? Does a new drug result in a lower mortality rate? These are just two examples of the kinds of questions that can be addressed with A/B testing, a procedure popular not only in business and medical clinical trials, but also in fields such as psychology, neuroscience, and biology. The A/B test set-up discussed in this article assumes that the outcome variable is binary; nevertheless, the outcome variable could in principle also be continuous. Based on a binary outcome variable, an A/B test compares the success rate of two options or treatment arms, A and B, and therefore can be conceptualized as a test for a difference between two proportions \citep{Little1989}. Typically, options A and B correspond to a control condition and an intervention or treatment of interest. 

Regardless of the specific field of application, we believe three general desiderata for A/B tests can be identified. First, we believe it is desirable that evidence can be obtained in favor of the null hypothesis that there is no difference between options A and B.
For instance, suppose a programmer alters code that should leave the appearance of a website unaffected. An A/B test may be conducted to confirm that the code changes did not lead to unintended consequences. Alternatively, suppose that a cheaper drug is introduced as a replacement of the standard drug; here, an A/B test may confirm that the cheaper drug is as effective as the drug that is currently standard.

Second, we believe it is desirable that evidence can be monitored as the data accumulate. Data collection can be time-consuming and expensive, and interim tests allow one to assess whether the results in hand are already sufficiently compelling or whether additional data ought to be obtained. There is also an ethical aspect to this desideratum, one that is particularly pronounced in case of new clinical treatments that are potentially beneficial or harmful; it is unethical to withhold treatment that interim analysis shows to be beneficial, just as it is unethical to continue to administer a treatment that interim analysis shows to be harmful (e.g., \citealp{Armitage1960}; see also \citealp{Ware1989} and the accompanying discussion). 

Third, we believe it is desirable that expert knowledge can be taken into account \citep[e.g.,][]{OHagan20019}.
In many A/B testing applications, there exists considerable expert knowledge about what size of effect to expect. For instance, the effect of website changes on conversion rates is often less than 0.5\% \citep{BermanEtAl2018}. Incorporating such expert knowledge into the statistical analysis will yield a more targeted test. 

The majority of A/B testing procedures that are currently in vogue do not fulfill the above desiderata. Specifically, many companies apply standard $p$-value-based null hypothesis significance testing to assess whether or not options A and B differ. This procedure has the advantage that it is readily available in software such as \proglang{R} \citep[e.g., via the functions \code{prop.test}, \code{fisher.test}, and \code{chisq.test}]{R}. However, this approach cannot distinguish between \emph{absence of evidence} (i.e., the data are inconclusive) and \emph{evidence of absence} (i.e., the data provide support for the null hypothesis that options A and B do not differ; \citealp[e.g.,][]{Dienes2014,KeysersEtAl2020}). Furthermore, although common practice, sequentially monitoring the uncorrected $p$-value (and stopping data collection as soon as the $p$-value is smaller than some fixed $\alpha$-level) invalidates the analysis (e.g., \citealp{Feller1940}). However, there exist valid classical sequential procedures that enable one to monitor a corrected $p$-value as data accumulate \citep[e.g.,][]{MalekEtAl2017}. For instance, \emph{Optimizely}, one of the leading commercial A/B testing platforms, has recently implemented an alternative $p$-value-based approach that allows users to continuously monitor the test outcome \citep{JohariEtAl2017}. Nevertheless, these sequential $p$-value-based procedures retain the inability to quantify evidence for the absence of an effect. Furthermore, (sequential) $p$-value-based A/B testing does not allow one to incorporate expert knowledge into the statistical analysis in a straightforward manner.

An alternative A/B testing approach that has become more popular of late is Bayesian estimation. For instance, \emph{VWO}, another leading A/B testing platform, has recently implemented a Bayesian estimation approach \citep{Stucchio2015}. A Bayesian estimation approach is also available via the \pkg{BayesianFirstAid} package \citep{Baath2014BayesianFirstAid} and the \pkg{bayesAB} package \citep{bayesAB}.\footnote{The \pkg{bayesAB} package provides a range of functions for Bayesian A/B testing. One advantage is that users can choose from a range of different data distributions (e.g., Bernoulli, normal, Poisson, etc.).} Since Bayesian inference does not require sample sizes to be fixed a priori \citep{BergerWolpert1988}, this approach allows one to monitor the analysis output as data accumulate. A Bayesian estimation approach also enables the incorporation of expert knowledge via the specification of a prior distribution that captures the expert's knowledge about a parameter of interest. However, this approach operates under the assumption that an effect exists --since a continuous prior assigns zero probability to a single null value-- and consequently does not allow one to obtain evidence in favor of the null hypothesis of no effect. For instance, \pkg{bayesAB} and \pkg{BayesianFirstAid} provide the user with the posterior probability that one option yields more successes than the other, but this ignores the fact that both options could be equally effective. Furthermore, the currently used Bayesian estimation approaches --such as the one implemented in \pkg{bayesAB} and \pkg{BayesianFirstAid}-- typically assign independent priors to the success probabilities of the control and treatment condition, a practice that was critiqued by \citet{Howard1998}.\footnote{``do English or Scots cattle have a higher proportion of cows infected with a certain virus? Suppose we were informed (before collecting any data) that the proportion of English cows infected was $0.8$. With independent uniform priors we would now give $H_1$ ($p_1 > p_2$) a probability of $0.8$ (because the chance that $p_2 > 0.8$ is still $0.2$). In very many cases this would not be appropriate. Often we will believe (for example) that if $p_1$ is 80\%, $p_2$ will be near 80\% as well and will be almost equally likely to be larger or smaller.'' (p. 363)}

To overcome the limitations of the current A/B tests we developed the \pkg{abtest} package in \proglang{R} \citep{R}. The \pkg{abtest} package implements one form of Bayesian inference for the A/B test, using informed prior distributions that induce a dependency between the two success probabilities. The analysis approach is based on a model by \citet{KassVaidyanathan1992}; for alternative approaches see \citet{DengEtAl2016}, \citet{JamilEtAl2017}, \citet{PhamEtAl2017}, and \citet{Skorski2019}. The implemented Bayesian procedure allows users (1) to obtain evidence in favor of the null hypothesis \citep[e.g.,][]{BergerDelampady1987,WagenmakersEtAl2018PBRPartI}; (2) monitor the evidence as the data accumulate \citep[e.g.,][]{Rouder2014PBR}; and (3) elicit and incorporate expert prior knowledge \citep[e.g.,][]{OHagan20019}. The \pkg{abtest} package thus fulfills all three desiderata mentioned above.

The \pkg{abtest} package provides functionality for both hypothesis testing and parameter estimation. In line with \citet{Jeffreys1939} and \citet{Fisher1928}, we believe that testing and estimation are complementary activities \citep{HaafEtAl2019}: before a parameter is estimated, it should be tested whether there is anything to justify estimation at all. \citet[p. 345]{Jeffreys1939} related this principle to Occam's razor: ``variation must be taken as random until there is positive evidence to the contrary'' \citep[see also][Section 8.1]{KassRaftery1995}. However, some researchers and practitioners oppose this idea, for instance because they believe that one should replace hypothesis testing with parameter estimation (\citealp[e.g.,][]{GelmanRubin1995}; \citealp{Cumming2014}). Nevertheless, the \pkg{abtest} package may also be useful for researchers without an interest in hypothesis testing, since the package can also be used exclusively for Bayesian parameter estimation (and prior elicitation).

This article is organized as follows: The next section introduces a fictitious business example. Afterwards, the implementation details of the Bayesian A/B test procedure used in \pkg{abtest} are discussed. Subsequently, the fictitious example is continued and the functionality of the \pkg{abtest} package and the practical benefits of the implemented approach are demonstrated. Next, a real data medical example is used to demonstrate further functionality of the package. The article ends with concluding comments.

\section{Example 1: effectiveness of resilience training}
Suppose the managers of a large consultancy firm are interested in reducing the number of employees who quit within the first six months, possibly due to the high stress involved in the job. A coaching company offers a resilience training and claims that this training greatly reduces the number of employees who quit. Implementing the training for all newly hired employees would be expensive and some of the managers are not completely convinced that the training is at all effective. Therefore, the managers decide to run an A/B test where half of a sample of newly hired employees will receive the training, the other half will not be trained. The outcome variable is whether or not an employee quit within the first six months (1 = still on the job, 0 = quit).

\begin{figure}
\centering
    \includegraphics[width = \textwidth]{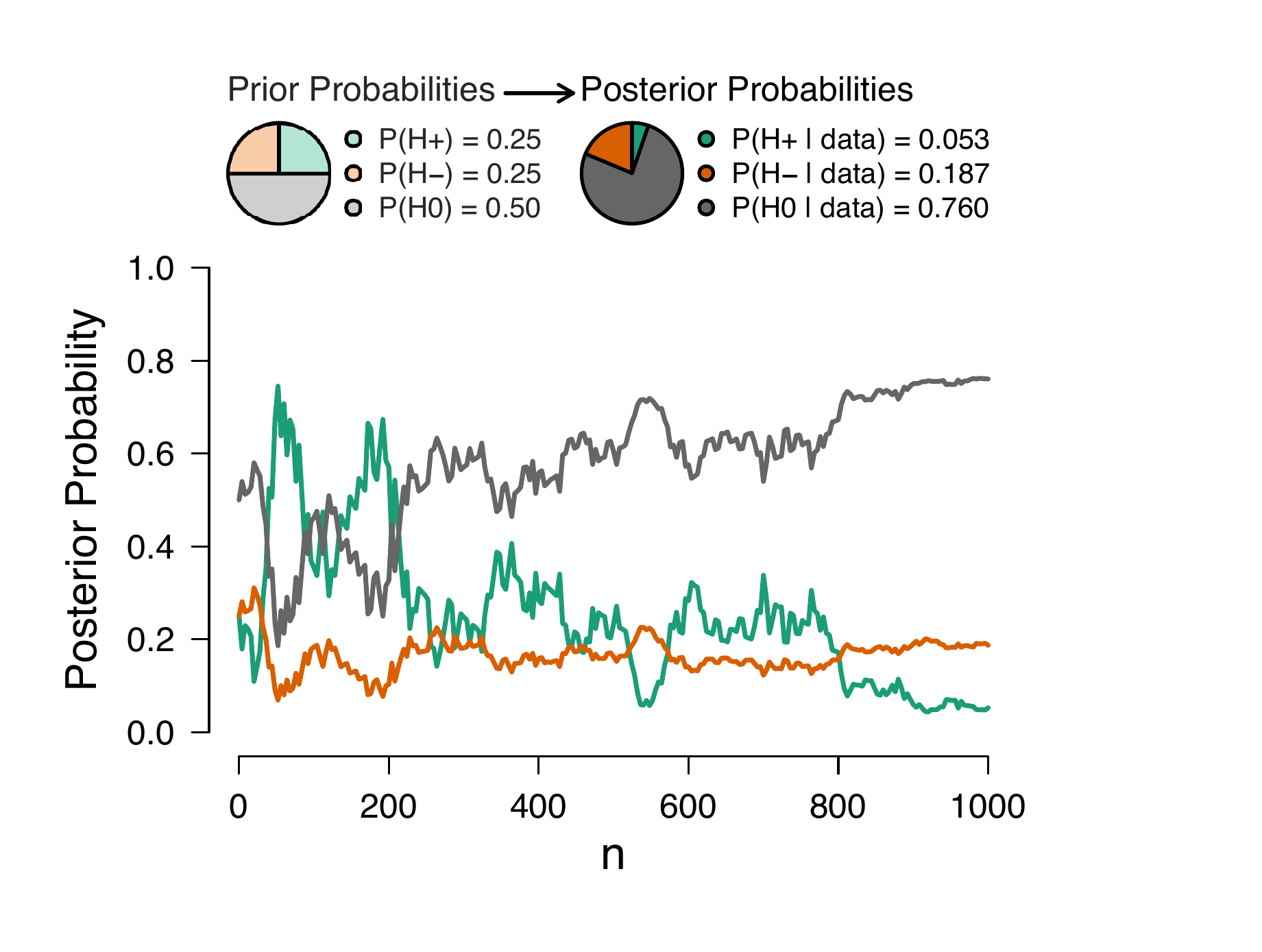}
	\caption{The posterior probability of the hypothesis that the training in Example~1 has a positive effect (i.e., $\mathcal{H}_+$), negative effect (i.e., $\mathcal{H}_-$), and no effect (i.e., $\mathcal{H}_0$) is plotted as a function of the number of observations across groups. On top, two probability wheels visualize the prior probabilities of the hypotheses and the posterior probabilities after taking into account all observations.}
	\label{fig:sequential}
\end{figure}

The consultancy firm collects $1,000$ observations ($500$ in each group). These (fictitious) data\footnote{The data set is structured such that the sequential nature of the data is retained: the data set contains the number of observations and the number of successes in each of the two groups after each observation.} are included in the \pkg{abtest} package (i.e., \code{seqdata}). The number of employees still on the job after six months is $249$ in the group without training and $269$ in the trained group. 
Figure~\ref{fig:sequential} provides an illustration of some of the information that can be obtained by analyzing these data using \pkg{abtest}. The figure displays the probability of the hypothesis that the training has a positive effect (i.e., $\mathcal{H}_+$), negative effect (i.e., $\mathcal{H}_-$), and no effect (i.e., $\mathcal{H}_0$) as a function of the number of observations across the two groups. The top part of the figure displays the probability of the three hypotheses before and after taking into account the observed data (i.e., prior and posterior probabilities) as probability wheels \citep[e.g.,][]{Tversky1969,LipkusHollands1999}.
Before providing more details about how to obtain and interpret this result as well as providing additional analyses, we discuss the implementation details of the A/B test procedure used by \pkg{abtest}.

\section{Implementation details}
The Bayesian A/B test implemented in the \pkg{abtest} package is based on \citet[Section 3, ``Testing Equality of Two Binomial Proportions'']{KassVaidyanathan1992}. Appendix A-C provide detailed derivations.

\subsection{Model}
Let $y_1$ denote the number of successes for option A with $n_1$ denoting the corresponding total number of observations for option A. Similarly, $y_2$ denotes the number of successes for option B with $n_2$ denoting the corresponding total number of observations for option B.
The Bayesian A/B test model based on \citet{KassVaidyanathan1992} is specified as follows:\footnote{Note that this is equivalent to a logistic regression model with a binary covariate (i.e., group membership) that is coded using $\pm 0.5$.}
\begin{equation}
\begin{split}
&\log\left(\frac{p_1}{1 - p_1}\right) = \beta - \frac{\psi}{2} \\
&\log\left(\frac{p_2}{1 - p_2}\right) = \beta + \frac{\psi}{2} \\
&y_1 \sim \text{Binomial}(n_1, p_1) \\
&y_2\sim \text{Binomial}(n_2, p_2).
\end{split}
\label{eq:model}
\end{equation}
Therefore, the model assumes that $y_1$ and $y_2$ follow binomial distributions with success probabilities $p_1$ and $p_2$. These probabilities are functions of the two model parameters, $\beta$ and $\psi$. Specifically, the log odds corresponding to $p_1$ are given by $\beta - \psi / 2$ and the log odds corresponding to $p_2$ are given by $\beta + \psi / 2$. 
The nuisance parameter $\beta$ corresponds to the grand mean of the log odds and the test-relevant parameter $\psi$ corresponds to the log odds ratio.
When $\psi$ is positive, this implies that $p_2 > p_1$ (i.e., option B has a higher success probability than option A); when $\psi$ is negative this implies that $p_2 < p_1$ (i.e., option B has a lower success probability than option A).

\subsection{Hypotheses}
The \pkg{abtest} package enables both estimation of the model parameters and testing of hypotheses about the test-relevant log odds ratio parameter $\psi$. There are four hypotheses that are of potential interest:
\begin{enumerate}
\item The null hypothesis $\mathcal{H}_0$ which states that the success probabilities $p_1$ and $p_2$ are identical, that is, $p_1 = p_2$. This is equivalent to $\mathcal{H}_0: \psi = 0$. This hypothesis corresponds to the claim that there is no difference between options A and B (i.e., the ``A/A test'').
\item The two-sided alternative hypothesis $\mathcal{H}_1$ which states that the two success probabilities $p_1$ and $p_2$ are not equal (i.e., $p_1 \ne p_2$), but does not specify which of the two is larger. This is equivalent to $\mathcal{H}_1: \psi \ne 0$. This hypothesis corresponds to the claim that options A and B differ but it is not specified which one yields more successes.
\item The one-sided hypothesis $\mathcal{H}_+$ which states that the second success probability $p_2$ is larger than the first success probability $p_1$. This is equivalent to $\mathcal{H}_+: \psi > 0$. This hypothesis corresponds to the claim that option B yields more successes than option A.
\item The one-sided hypothesis $\mathcal{H}_-$ which states that the first success probability $p_1$ is larger than the second success probability $p_2$. This is equivalent to $\mathcal{H}_-: \psi < 0$. This hypothesis corresponds to the claim that option A yields more successes than option B. 
\end{enumerate}

Researchers who conduct an A/B test are usually interested in answering the question: Does option B yield more successes than option A (i.e., $\mathcal{H}_+$), fewer successes than option A (i.e., $\mathcal{H}_-$), or is there no difference between options A and B (i.e., $\mathcal{H}_0$)? Therefore, it may be argued that the hypotheses of interest are typically $\mathcal{H}_+$, $\mathcal{H}_-$, and $\mathcal{H}_0$. Consequently, by default, only these three hypotheses are assigned non-zero prior probability in the \pkg{abtest} package. Specifically, a default prior probability of $.50$ is assigned to the hypothesis that there is no effect (i.e., $\mathcal{H}_0$), and the remaining prior probability is split evenly across the hypothesis that there is a positive effect (i.e., $\mathcal{H}_+$ receives $.25$) and a negative effect (i.e., $\mathcal{H}_-$ also receives $.25$). The user may change these default prior probabilities to custom values.
\begin{table}[!tb]
	\centering
	\caption{Changing the prior probability assignments across rival hypotheses produces different tests.}
	\begin{tabular}{cccccc}
	\toprule
	& \multicolumn{5}{c}{Test}\\
	\cmidrule(lr){2-6}
  Hypothesis & Default & Undirected & Positive & Negative & Direction \\
\cmidrule(lr){1-1} \cmidrule(lr){2-6}
$\mathcal{H}_0$ & \textbf{.50} & \textbf{.50} & \textbf{.50} & \textbf{.50} & 0\\

$\mathcal{H}_1$ & 0 & \textbf{.50} & 0 & 0 & 0\\

$\mathcal{H}_+$ & \textbf{.25} & 0 & \textbf{.50} & 0 & \textbf{.50}\\

$\mathcal{H}_-$ & \textbf{.25} & 0 & 0 & \textbf{.50} & \textbf{.50}\\
\bottomrule
\end{tabular}
\label{tab:prior_prob}
\end{table}
Table~\ref{tab:prior_prob} provides an overview of five qualitatively different tests that can be conducted by assigning prior probabilities to hypotheses in certain ways.\footnote{Note that, except for the first column of Table~\ref{tab:prior_prob} which displays the default setting, the remaining examples use equal prior probabilities for all hypotheses that are assigned non-zero prior probability. However, the user can of course also assign prior probability unevenly to the hypotheses of interest (e.g., if prior knowledge exists about the relative plausibility of the rival hypotheses).}
The first column displays the default setting that assigns probability $.50$ to the null hypothesis and splits the remaining probability evenly across $\mathcal{H}_+$ and $\mathcal{H}_-$.
The second column displays a prior probability assignment that implements an undirected test (i.e., $\mathcal{H}_0$ is compared to the undirected $\mathcal{H}_1$).
The third column displays a prior probability assignment for testing whether the effect is non-existent or positive. The fourth column displays a prior probability assignment for testing whether the effect is non-existent or negative. Finally, the fifth column displays a prior probability assignment for a test of direction, that is, for testing whether the effect is positive or negative. This last setting may be of interest whenever the null hypothesis is a priori deemed implausible, uninteresting, or irrelevant. 

\subsection{Parameter priors}
The \pkg{abtest} package assigns normal priors to the model parameters: $\beta \sim \mathcal{N}(\mu_\beta, \sigma_\beta^2)$ and $\psi \sim \mathcal{N}(\mu_\psi, \sigma_\psi^2)$. As illustrated in the example below, these priors result in a dependency in the implied prior for the success probabilities $p_1$ and $p_2$, which is generally desirable \citep{Howard1998}.

For the one-sided hypotheses $\mathcal{H}_+$ and $\mathcal{H}_-$, the prior on $\psi$ is truncated at zero. Specifically, for $\mathcal{H}_+$, the prior on $\psi$ is a truncated normal distribution with parameters $\mu_\psi$ and $\sigma_\psi$ and lower bound at zero. For $\mathcal{H}_-$, the prior on $\psi$ is a truncated normal distribution with parameters $\mu_\psi$ and $\sigma_\psi$ and upper bound at zero.
These normal priors are computationally convenient and sufficiently flexible to encode a wide range of prior information.

By default, the \pkg{abtest} package assigns standard normal priors to both $\beta$ and $\psi$.
For the nuisance parameter $\beta$, a standard normal prior results in  a relatively flat implied prior on $p_1$ and $p_2$ when $\psi = 0$. Generally, the choice of a prior for the nuisance parameter $\beta$ is relatively inconsequential \citep{KassVaidyanathan1992}. In contrast, the prior on the test-relevant parameter $\psi$ is consequential, as it defines the extent to which the hypotheses of interest differ from $\mathcal{H}_0$. Our choice for a default standard normal prior on the test-relevant parameter $\psi$ is motivated by the fact that a zero-centered prior does not favor any of the two options A or B a priori. Furthermore, the standard deviation of 1 results in a prior distribution that assigns mass to a wide range of reasonable log odds ratios \citep{ChenEtAl2010} without being so uninformative that the results unduly favor $\mathcal{H}_0$ \citep{Bartlett1957,Lindley1957}.\footnote{Note that the default implied prior on the absolute risk $p_2-p_1$ is considerably more narrow than the prior induced by the popular default choice that assigns $p_1$ and $p_2$ independent uniform distributions \citep{Jeffreys1935}.} However, large changes in the prior standard deviation of the test-relevant parameter may result in large changes in the results, as the prior standard deviation governs the degree to which the hypothesis of interest makes predictions that differ from $\mathcal{H}_0$. To include prior knowledge about the expected results, the \pkg{abtest} package allows the user to change the default values of the prior distributions for the nuisance parameter $\beta$ and the test-relevant parameter $\psi$, either by changing the location of the normal prior distribution, the scale, or both.

\subsection{Encoding prior information}
A straightforward way to encode prior information about the model parameters is to set $\mu_\beta$, $\sigma_\beta$, $\mu_\psi$, and $\sigma_\psi$ directly.
However, it may sometimes be easier to specify prior distributions based on quantities such as the (log) odds ratio, relative risk (i.e., $p_2/p_1$, the ratio of the success probability in condition B and condition A), and absolute risk (i.e., $p_2 - p_1$, the difference of the success probability in condition B and condition A).
The \code{elicit_prior} function allows users to encode prior information about a quantity of interest (either log odds ratio, odds ratio, relative risk, or absolute risk). The function assumes that the prior on $\beta$ is not the primary target of prior elicitation and is fixed by the user a priori (using the arguments \code{mu_beta} and \code{sigma_beta}) -- for instance, to a standard normal prior which corresponds to a relatively flat implied prior on $p_1$ and $p_2$ when $\psi = 0$. 

To encode prior information, the user needs to provide quantiles for a quantity of interest. Let $q_i, i = 1,\ldots,I$ denote the values of $I$ quantiles provided by the user and let $\text{prob}_i, i = 1,\ldots,I$ denote the corresponding probabilities  (e.g., for the median, $\text{prob}_i = 0.5$).
Least-squares minimization is used to obtain $\mu_\psi$ and $\sigma_\psi$ as follows:
\begin{equation}
(\mu_\psi, \sigma_\psi) = \underset{\mu_\psi, \sigma_\psi}{\mathrm{arg \, min}} \sum_{i = 1}^{I} \left(F(q_i; \mu_\psi, \sigma_\psi) - \text{prob}_i\right)^2,
\label{eq:elicitation}
\end{equation}
where $F(\cdot; \mu_\psi, \sigma_\psi)$ corresponds to the cumulative distribution function (cdf) for the quantity of interest implied by the normal prior on $\psi$. For some quantities, this cdf also depends on the prior for $\beta$; however, as described above, it is assumed that $\mu_\beta$ and $\sigma_\beta$ are fixed a priori.

\subsection{Hypothesis testing}
To quantify the evidence that the data provide for $\mathcal{H}_0$, $\mathcal{H}_1$, $\mathcal{H}_+$, and $\mathcal{H}_-$, one can compute Bayes factors \citep{Jeffreys1939,KassRaftery1995} and posterior probabilities of the rival hypotheses.
The posterior probability of hypothesis $\mathcal{H}_j$, $j \in \{0, 1, +, -\}$ is given by:
	\begin{equation}
	\label{eq:post_model_probs}
	\overbrace{p(\mathcal{H}_j \mid \text{data})}^{\text{posterior probability}} = \overbrace{\frac{p(\text{data} \mid \mathcal{H}_j)}{\sum_{k} p(\text{data} \mid \mathcal{H}_k) \, p(\mathcal{H}_k)}}^{\text{updating factor}} \;\;\;\;\;\; \times \overbrace{p(\mathcal{H}_j)}^{\text{prior probability}}.
	\end{equation}
The Bayes factor for comparing hypotheses $\mathcal{H}_j$ and $\mathcal{H}_k$ equals the change from prior to posterior odds:
\begin{equation}
	\label{eq:post_model_odds}
	\underbrace{\frac{p(\mathcal{H}_j \mid \text{data})}{p(\mathcal{H}_k \mid \text{data})}}_{\text{posterior odds}} = \underbrace{\frac{p(\text{data} \mid \mathcal{H}_j)}{p(\text{data} \mid \mathcal{H}_k)}}_{\text{Bayes factor BF$_{jk}$}} \times \underbrace{\frac{p(\mathcal{H}_j)}{p(\mathcal{H}_k)}}_{\text{prior odds}}.
	\end{equation}
In order to obtain posterior probabilities of the hypotheses and Bayes factors one needs to evaluate the marginal likelihood $p(\text{data} \mid \mathcal{H}_j)$ for each hypothesis $j \in \{0, 1, +, -\}$. For $\mathcal{H}_0$ and $\mathcal{H}_1$, we evaluate the marginal likelihood using Laplace approximations as suggested by \citet{KassVaidyanathan1992}.
Specifically, the marginal likelihood for $\mathcal{H}_0$ is approximated by:
\begin{equation}
\begin{split}
p(\text{data} \mid \mathcal{H}_0) &=  \int \underbrace{p(\text{data} \mid \beta)}_{\text{likelihood}} \, \underbrace{\pi_0(\beta)}_{\text{prior}} \text{d}\beta \\
&\approx (2 \pi \sigma_0^2)^\frac{1}{2} \, \exp\left\{l_0^\ast(\beta_0^\ast)\right\},
\end{split}
\end{equation}
where $l_0^\ast(\beta) = \log\left\{p(\text{data} \mid \beta) \, \pi_0(\beta)\right\}$, $\beta_0^\ast$ corresponds to the mode of $l_0^\ast(\beta)$, and \sloppy$\sigma_0^2 = \left(-\frac{d^2}{d\beta^2} \, l_0^\ast(\beta)\right)^{-1} \bigg\rvert_{\beta = \beta_0^\ast}$ denotes the inverse of the negative second derivative of $l_0^\ast(\beta)$ evaluated at the mode $\beta_0^\ast$.

The marginal likelihood for $\mathcal{H}_1$ is approximated by:
\begin{equation}
\begin{split}
p(\text{data} \mid \mathcal{H}_1) &=  \int \int \underbrace{p(\text{data} \mid \beta, \psi)}_{\text{likelihood}} \, \underbrace{\pi(\beta, \psi)}_{\text{prior}} \text{d}\beta \text{d}\psi\\
&\approx 2 \pi \, \det\left(\boldsymbol{\Sigma}_1\right)^\frac{1}{2} \exp\left\{l^\ast(\beta^\ast, \psi^\ast)\right\},
\end{split}
\end{equation}
where $l^\ast(\beta, \psi) = \log\left\{p(\text{data} \mid \beta, \psi) \, \pi(\beta, \psi)\right\}$, $(\beta^\ast, \psi^\ast)$ denotes the mode of $l^\ast(\beta, \psi)$, and \sloppy$\boldsymbol{\Sigma}_1 = \left(- \boldsymbol{H}_1\right)^{-1}\big\rvert_{(\beta, \psi) = (\beta^\ast, \psi^\ast)}$ denotes the inverse of the negative Hessian $\boldsymbol{H}_1$ (i.e., the matrix with second-order partial derivatives) of $l^\ast(\beta, \psi)$ evaluated at the mode $(\beta^\ast, \psi^\ast)$.

\begin{figure}
	\includegraphics[width = \textwidth]{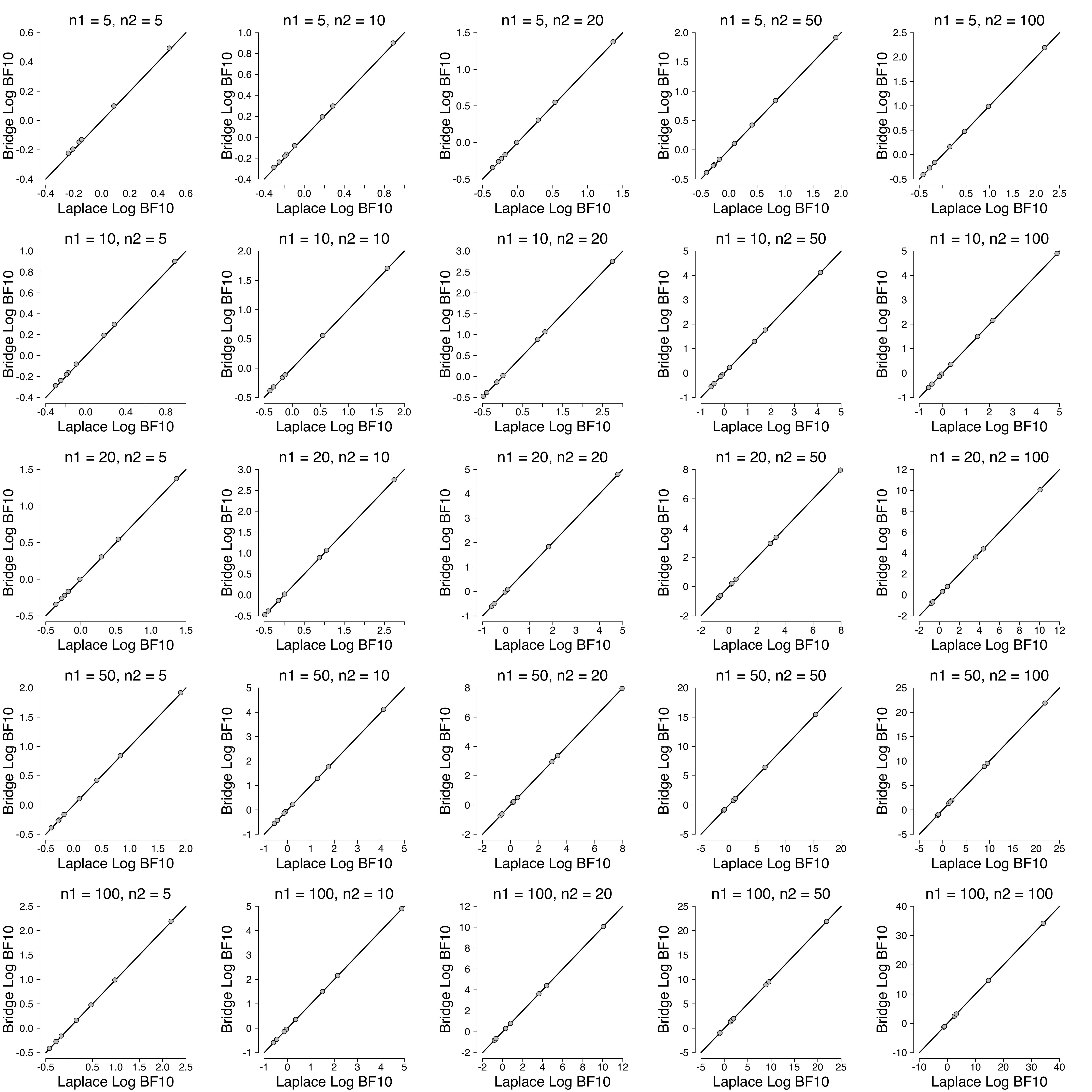}
	\caption{Comparison of the Laplace approximation and bridge sampling for computing the (log of the) Bayes factor $\text{BF}_{10}$. We considered all possible combinations of $n_1 \in \{5, 10, 20, 50, 100\}$ and $n_2 \in \{5, 10, 20, 50, 100\}$. For each of the $n_1$-$n_2$ combinations, we considered all possible combinations of $y_1 \in \{\frac{1}{5} n_1, \frac{2}{5}  n_1, \frac{3}{5}  n_1, \frac{4}{5}  n_1\}$ and $y_2 \in \{\frac{1}{5}  n_2, \frac{2}{5}  n_2, \frac{3}{5}  n_2, \frac{4}{5}  n_2\}$. The results reveal that the two methods yield highly similar results, even when sample size is very small.}
	\label{fig:bridge_laplace}
\end{figure}

These Laplace approximations work well in practice, even for sample sizes that are extremely small.
As a demonstration, for a range of synthetic data sets we computed the (log of the) Bayes factor $\text{BF}_{10}$ which compares $\mathcal{H}_1$ to $\mathcal{H}_0$ using the above Laplace approximations and, as a comparison, also using bridge sampling \citep{MengWong1996, GronauEtAlbridgesampling}. The priors on $\beta$ and $\psi$ were standard normal distributions. 
Figure~\ref{fig:bridge_laplace} displays the results and confirms that the Laplace approximation yields accurate results, even for sample sizes as small as $n_1 = n_2 = 5$.

For the one-sided hypotheses $\mathcal{H}_+$ and $\mathcal{H}_-$, Laplace approximations did not appear to yield accurate results for small sample sizes, even after removing the constraint on $\psi$ through the parameterization $(\beta, \xi) = (\beta, \log\left(\psi\right))$ for $\mathcal{H}_+$ and $(\beta, \xi) = (\beta, \log\left(- \psi\right))$ for $\mathcal{H}_-$.  The \pkg{abtest} package therefore uses importance sampling to increase the accuracy of the Laplace approximations when computing the marginal likelihoods for $\mathcal{H}_+$ and $\mathcal{H}_-$.
Specifically, a Laplace approximation is used to approximate the mode and covariance matrix of the posterior. The importance density is then given by a multivariate $t$ distribution with location set to the approximated posterior mode, scale matrix set to the approximated posterior covariance matrix, and five degrees of freedom (note that the user can change the degrees of freedom).
The marginal likelihood for $\mathcal{H}_+$ is then estimated as follows:
\begin{equation}
\begin{split}
p(\text{data} \mid \mathcal{H}_+) &= \int \int \underbrace{p(\text{data} \mid \beta, \xi)}_{\text{likelihood}} \, \underbrace{\pi_+(\beta, \xi)}_{\text{prior}}  \text{d}\beta \text{d}\xi\\
&\approx \frac{1}{S} \sum_{s = 1}^{S} \frac{p(\text{data} \mid \tilde{\beta}_s, \tilde{\xi}_s) \, \pi_+(\tilde{\beta}_s, \tilde{\xi}_s)}{g_{\text{is}}(\tilde{\beta}_s, \tilde{\xi}_s)},
\end{split}
\end{equation}
where $\left\{\tilde{\beta}_s, \tilde{\xi}_s\right\}_{s = 1}^S$ denotes $S$ samples from the multivariate $t$ importance density $g_{\text{is}}$, and 
\begin{equation}
\pi_+(\beta, \xi) = \mathcal{N}(\beta; \mu_\beta, \sigma_\beta^2) \, \mathcal{N}_+(\exp(\xi); \mu_\psi, \sigma_\psi^2) \, \exp(\xi),
\end{equation}
where $\mathcal{N}(x; y, z)$ denotes the probability density function of a normal distribution with mean $y$ and variance $z$ that is evaluated at $x$. Furthermore, $\mathcal{N}_+(x; y, z)$ denotes the density of a normal distribution that is truncated to allow only positive values for $x$. The marginal likelihood for $\mathcal{H}_-$ is computed analogously.

\subsection{Obtaining posterior samples}
In a Bayesian A/B test application, one may not only be interested in testing hypotheses, but also in obtaining posterior samples for the model parameters under $\mathcal{H}_1$, $\mathcal{H}_+$, and $\mathcal{H}_-$. The \pkg{abtest} package allows the user to obtain posterior samples using sampling importance resampling \citep[e.g.,][]{RobertCasella2010}.
Specifically, posterior samples for $\mathcal{H}_+$ are obtained as follows (samples for the other hypotheses are obtained in an analogous manner):
\begin{enumerate}
\item Generate $S$ samples from the multivariate $t$ proposal distribution mentioned before, denoted by $\left\{\tilde{\beta}_s, \tilde{\xi}_s\right\}_{s = 1}^S$.
\item Compute the importance weights:
\begin{equation}
w_s = \frac{p(\text{data} \mid \tilde{\beta}_s, \tilde{\xi}_s) \, \pi_+(\tilde{\beta}_s, \tilde{\xi}_s)}{g_{\text{is}}(\tilde{\beta}_s, \tilde{\xi}_s)} \,\, , \hspace{2em} s = 1, 2, \ldots, S.
\end{equation}
\item Renormalize the importance weights: $v_s = w_s/\sum_{t = 1}^{S} w_t$, $s = 1, 2, \ldots, S$.
\item Resample (with replacement) from the samples obtained from the importance density according to the normalized importance weights $v_s$ which yields (approximate) samples from the posterior distribution.
\end{enumerate}

\section{Example 1: effectiveness of resilience training (continued)}

Next we continue the effectiveness of resilience training example and show how expert prior information can be taken into account, how the hypotheses of interest can be tested, and how one can estimate the model parameters using the \pkg{abtest} package.

\subsection{Prior specification}
\begin{figure}
\centering
    \begin{tabular}{c}
    \includegraphics[width = 0.6 \textwidth]{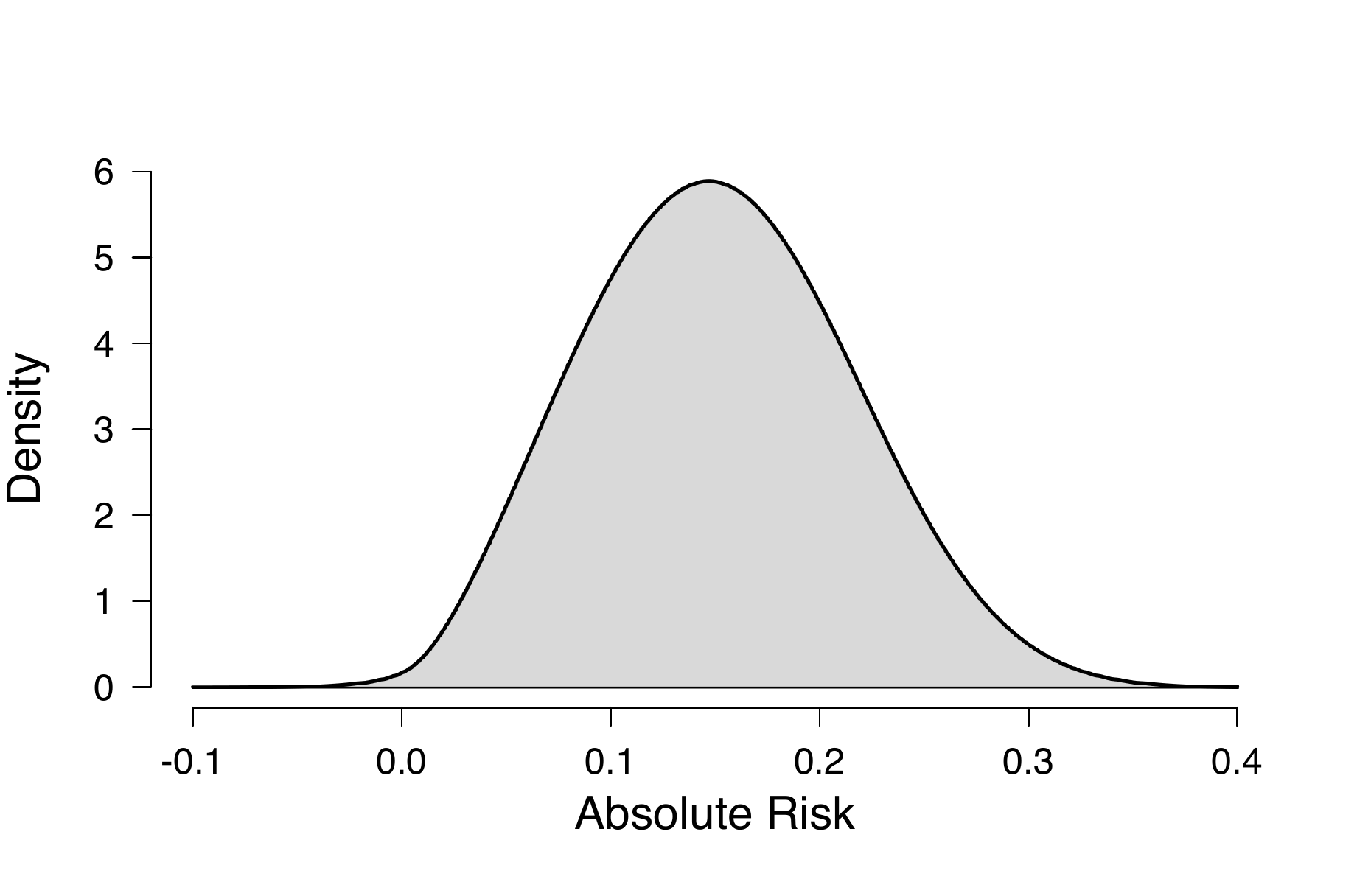} \\
    \includegraphics[width = 0.6 \textwidth]{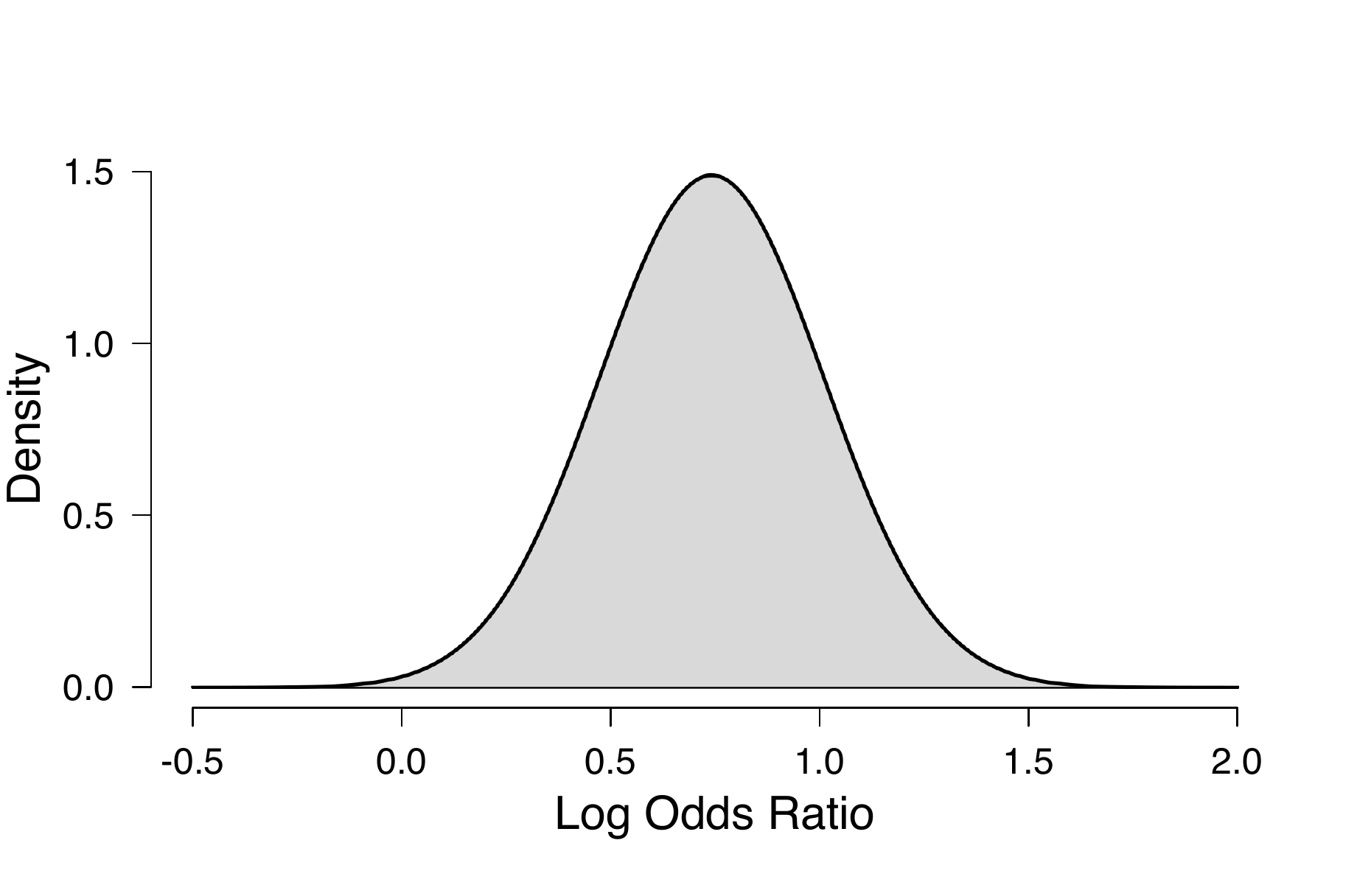} \\
    \includegraphics[width = 0.6 \textwidth]{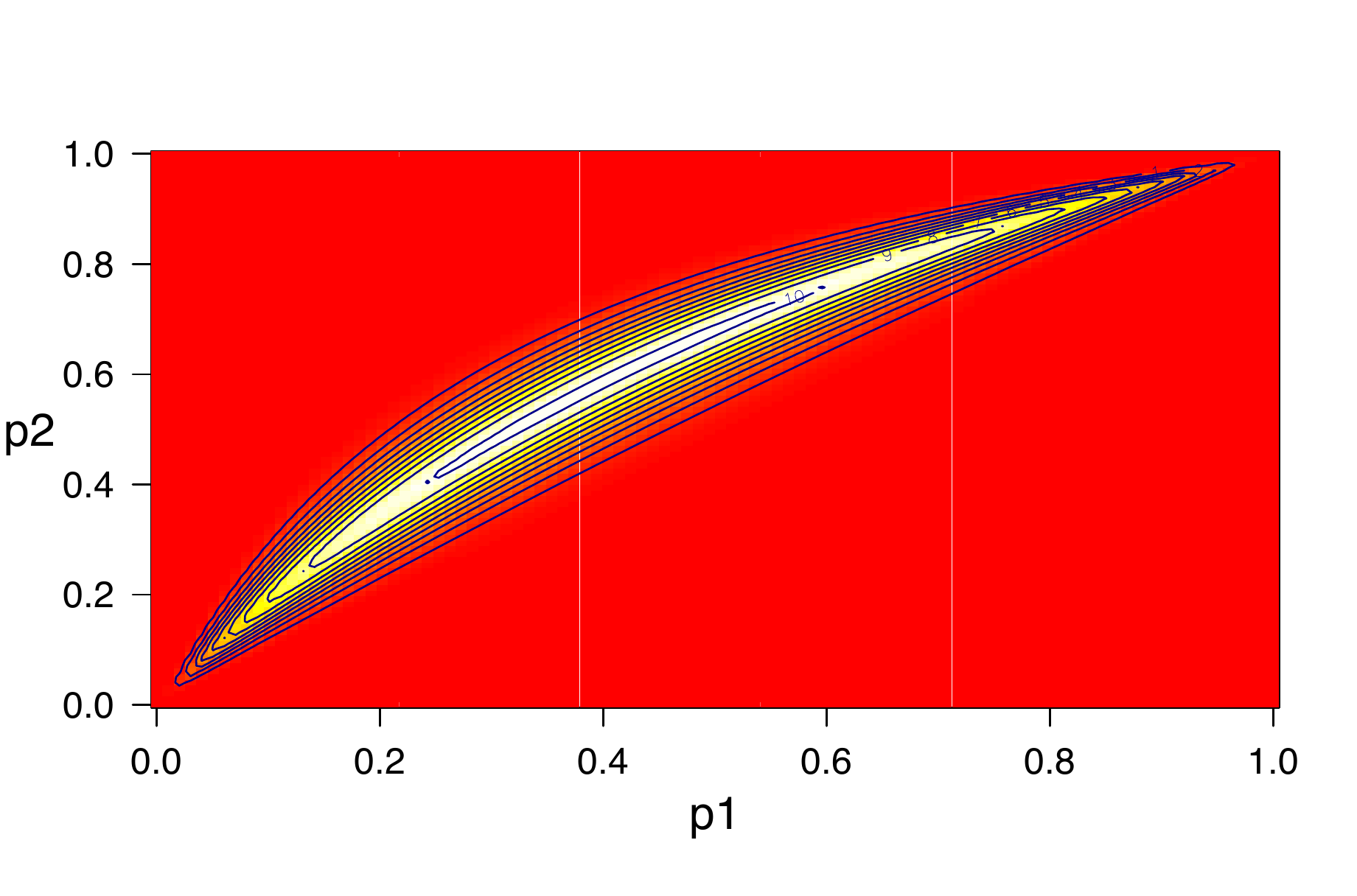}
    \end{tabular}
	\caption{Elicited (implied) prior distributions for the effectiveness of the resilience training in Example~1. The top panel displays the prior distribution for the absolute risk which corresponds to the difference between the probability of still being on the job for the trained and the non-trained employees (i.e., $p_2 - p_1$). The middle panel shows the prior distribution for the log odds ratio parameter $\psi$. The bottom panel displays the implied joint prior distribution for the success probabilities $p_1$ and $p_2$. The bottom panel illustrates that the two success probabilities are assigned dependent priors. Furthermore, most prior mass is above the main diagonal which represents the coaching company's prior expectation that the training is successful.}
	\label{fig:priors}
\end{figure}

Before commencing the A/B test, the managers asked the coaching company to specify how effective they believe the training will be. The coaching company claimed that, based on past experience with the training, they expect the proportion of employees who do not quit within the first six months to be 15\% larger for the group who received the training, with a 95\% uncertainty interval ranging from a 2.5\% benefit to a 27.5\% benefit. Assuming that the claimed 15\% corresponds to the prior median, this expectation corresponds to a median absolute risk (i.e., $p_2 - p_1$) of $0.15$ with a 95\% uncertainty interval ranging from $0.025$ to $0.275$. The \code{elicit_prior} function can be used to encode this prior information:\footnote{All code and plots are also available at \url{https://osf.io/t3ajr/}.}
\begin{Sinput}
R> library("abtest")
R> prior_par <- elicit_prior(q = c(0.025, 0.15, 0.275),
+                            prob = c(.025, .5, .975),
+                            what = "arisk")
\end{Sinput}
The obtained prior on the absolute risk can be visualized as follows:
\begin{Sinput}
R> plot_prior(prior_par, what = "arisk")
\end{Sinput}
The resulting graph is shown in the top panel of Figure~\ref{fig:priors}. 
The user can also visualize the (implied) prior for other quantities. For instance, the prior on the log odds ratio (middle panel of Figure~\ref{fig:priors}) is obtained as follows:
\begin{Sinput}
R> plot_prior(prior_par, what = "logor")
\end{Sinput}
The implied prior on the success probabilities $p_1$ and $p_2$ (bottom panel of Figure~\ref{fig:priors}) is obtained as follows:
\begin{Sinput}
R> plot_prior(prior_par, what = "p1p2")
\end{Sinput}
The bottom panel of Figure~\ref{fig:priors} illustrates that there is a dependency between $p_1$ and $p_2$ which is arguably desirable \citep{Howard1998}: When one of the success probabilities is very (small) large, it is likely that the other one will also be (small) large.

\subsection{Hypothesis testing}
Since the number of employees still on the job after six months is $249$ in the group without training and $269$ in the trained group, the observed success probabilities are $\hat{p}_1 = .498$ in the control group and $\hat{p}_2 = .538$ in the group that received training. Consequently, the observed success probabilities suggest that there is a positive effect of the training of 4\%; however, a statistical analysis is required to assess whether this observed difference is statistically compelling. The \code{ab_test} function can be used to conduct a Bayesian A/B test as follows:
\begin{Sinput}
R> data("seqdata")
R> set.seed(1)
R> ab <- ab_test(data = seqdata, prior_par = prior_par)
\end{Sinput}
This yields the following output:
\begin{Soutput}
R> print(ab)

Bayesian A/B Test Results:

 Bayes Factors:

 BF10: 0.1406443
 BF+0: 0.13823
 BF-0: 0.4920187

 Prior Probabilities Hypotheses:

 H+: 0.25
 H-: 0.25
 H0: 0.5

 Posterior Probabilities Hypotheses:

 H+: 0.0526
 H-: 0.1871
 H0: 0.7604
\end{Soutput}
The first part of the output presents Bayes factors in favor of the hypotheses $\mathcal{H}_1$, $\mathcal{H}_+$, and $\mathcal{H}_-$, where the reference hypothesis (i.e., denominator of the Bayes factor) is $\mathcal{H}_0$. Since all three Bayes factors are smaller than 1, they all indicate evidence in favor of the null hypothesis of no effect. The next part of the output displays the prior probabilities of the hypotheses with non-zero prior probability. As explained before, the default setting assigns probability $.50$ to the null hypothesis and splits the remaining probability evenly across $\mathcal{H}_+$ and $\mathcal{H}_-$. The user can change this default setting via the \code{prior_prob} argument (e.g., to assign non-zero probability to $\mathcal{H}_1$). The final part of the output displays the posterior probabilities of the hypotheses with non-zero prior probability. The posterior probability of the null hypothesis $\mathcal{H}_0$ indicates that the data have increased the plausibility of the null hypothesis from $.50$ to $.76$. Furthermore, the data have decreased the plausibility of both $\mathcal{H}_+$ and $\mathcal{H}_-$.

As an aside, it may appear paradoxical that the data indicate a 4\% positive effect of the training and yet the posterior probability of $\mathcal{H}_-$ is larger than that of $\mathcal{H}_+$. The reason for this result is that the company's prior was overly ambitious, and $\mathcal{H}_+$ is penalized for having predicted effects that are much too large. Furthermore, note that the test-relevant prior distribution under $\mathcal{H}_-$ is obtained by truncating the prior on $\psi$ at zero and renormalizing. Since the company's prior assigns almost all mass to positive log odds ratio values, renormalizing the negative part of the distribution results in a prior that is highly similar to $\mathcal{H}_0$; this explains why $\mathcal{H}_-$ receives non-trivial posterior probability. These considerations underscore the fact that the outcome of a Bayesian analysis is always relative to the specific set of models (and associated prior distributions) under consideration. Because highly informed priors can exert a large influence on the results, it is generally wise to examine the robustness of the conclusions by executing the default analysis as well. This analysis is reported in Appendix~D.

\begin{figure}
\centering
    \includegraphics[width = \textwidth]{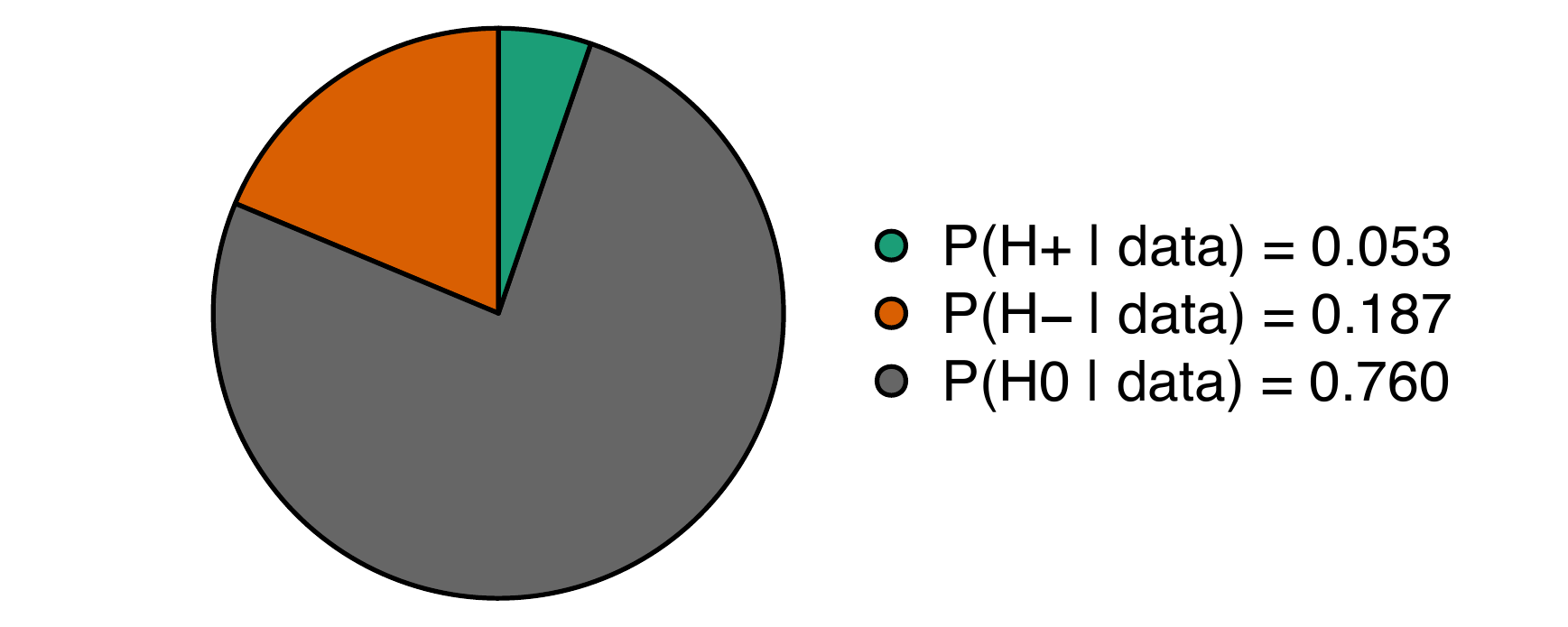}
	\caption{Posterior probabilities of the hypotheses visualized as a probability wheel for Example~1.}
	\label{fig:post_probs}
\end{figure}
The \pkg{abtest} package allows users to visualize the posterior probabilities of the hypotheses by means of a probability wheel (Figure~\ref{fig:post_probs}):
\begin{Sinput}
R> prob_wheel(ab)
\end{Sinput}
Overall, the data support the hypothesis that the training is ineffective over the company's hypothesis that the training is highly effective. The Bayes factor for $\mathcal{H}_0$ over $\mathcal{H}_+$ equals $1/0.138 \approx 7.2$, which indicates moderate evidence \citep[Appendix I]{Jeffreys1939}.   

Since the data set is of a sequential nature, it may be of interest to consider not only the result based on all observations, but to conduct also a sequential analysis that tracks the evidential flow as a function of the total number of observations (i.e., the number of observations across both groups). This sequential analysis can be conducted as follows:
\begin{Sinput}
R> plot_sequential(ab, thin = 4)
\end{Sinput}
Setting the \code{thin} argument to \code{4} indicates that the evidence is computed after every 4$th$ observation. Thinning can be useful to speed up the analysis in case the data set is very large or in case observations arrive in batches. Figure~\ref{fig:sequential} displays the result of the sequential analysis. The posterior probability of each hypothesis with non-zero prior probability is plotted as a function of the total number of observations. At the top, two probability wheels visualize the prior probabilities of the hypotheses and the posterior probabilities of the hypotheses based on all available data. Figure~\ref{fig:sequential} shows that after some initial fluctuation, adding more observations increased the probability of the null hypothesis that there is no effect of the training. 

\subsection{Parameter estimation}
The data indicate evidence in favor of the null hypothesis versus the hypothesis that the training is highly effective, leaving open the possibility that the training does have an effect, but of a more modest size than the company anticipated. To assess this possibility one may investigate the potential size of the effect under the assumption that the effect is non-zero.\footnote{For consistency, we continue this analysis with the company's prior; an analysis with the less enthusiastic default prior is provided in Appendix~D.} For parameter estimation, we generally prefer to investigate the posterior distribution for the unconstrained alternative hypothesis $\mathcal{H}_1$; however, the \pkg{abtest} package also provides posterior samples and plotting functionality for the constrained hypotheses $\mathcal{H}_+$ and $\mathcal{H}_-$.

\begin{figure}
\centering
    \begin{tabular}{c}
    \includegraphics[width = 0.56 \textwidth]{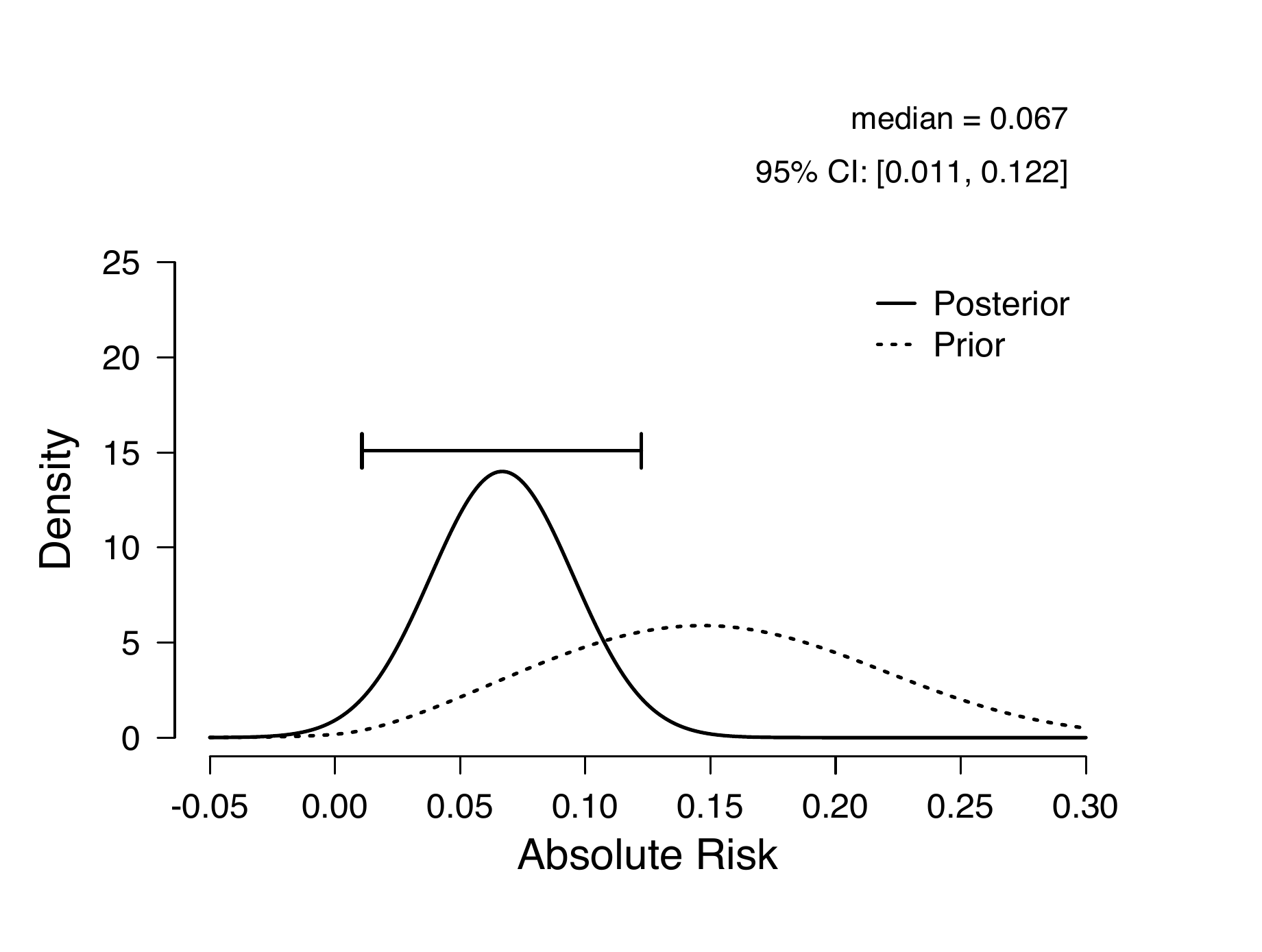} \\
    \includegraphics[width = 0.56 \textwidth]{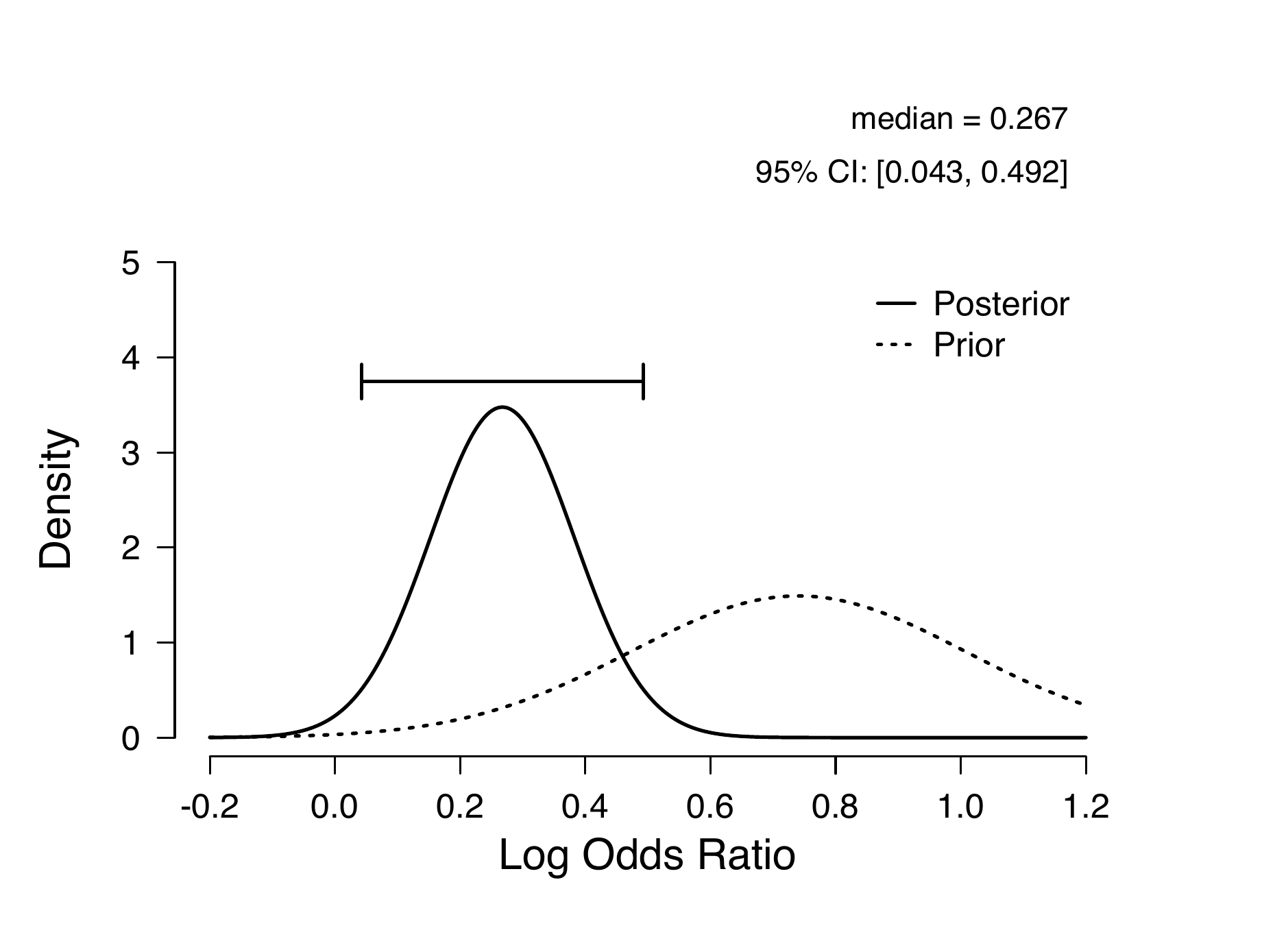} \\
    \includegraphics[width = 0.56 \textwidth]{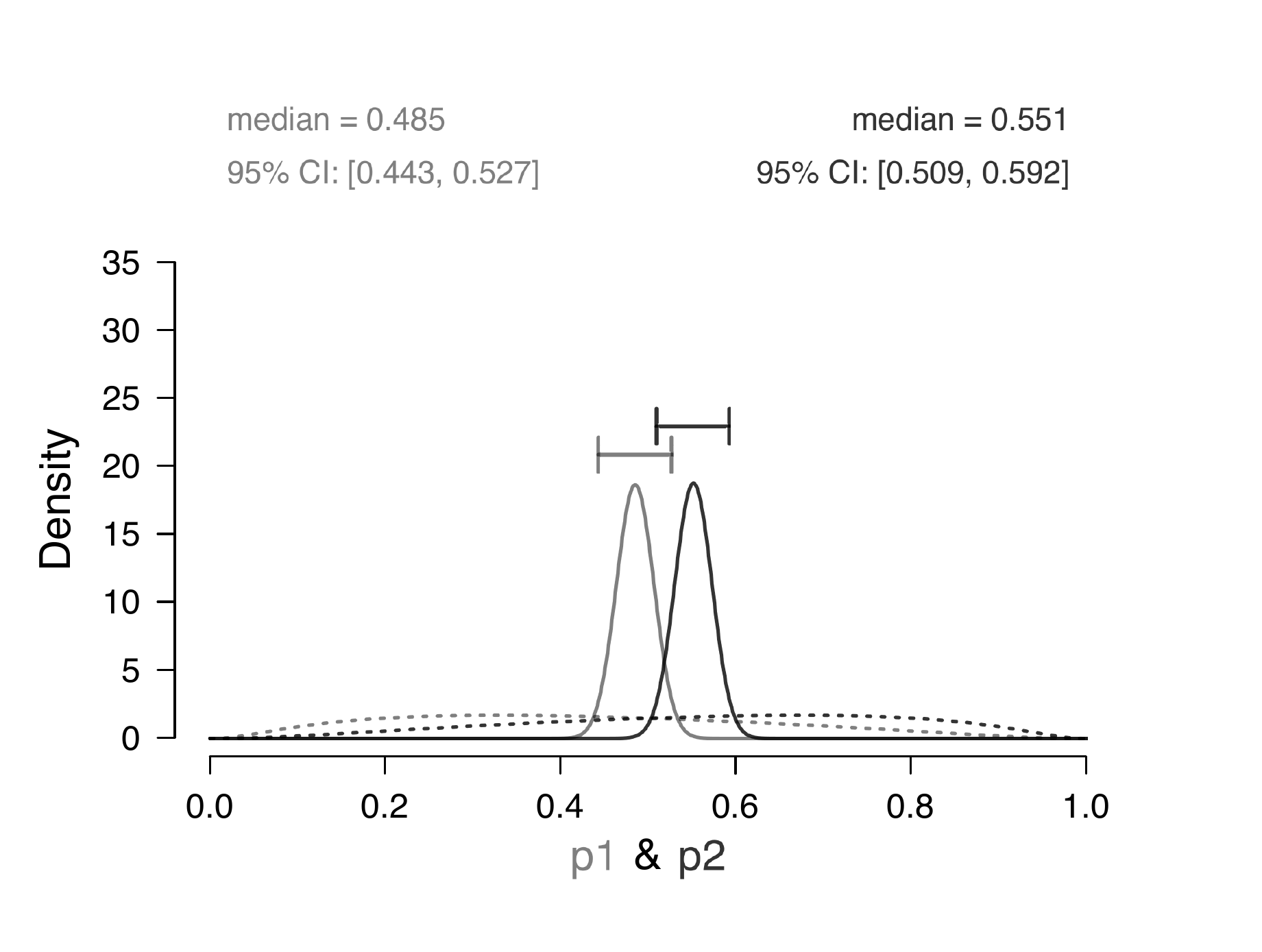}
    \end{tabular}
	\caption{(Implied) prior and posterior distributions under $\mathcal{H}_1$ for Example~1. The dotted lines display the prior distributions, the solid lines display the posterior distributions (with 95\% central credible intervals). The medians and the bounds of the 95\% central credible intervals are displayed on top of each panel. The top panel displays the posterior distribution for the absolute risk (i.e., $p_2 - p_1$); the middle panel shows the posterior distribution for the log odds ratio parameter $\psi$; the bottom panel displays the marginal posterior distributions for the success probabilities $p_1$ and $p_2$.}
	\label{fig:posteriors}
\end{figure}

The top panel of Figure~\ref{fig:posteriors} displays the posterior distribution for the absolute risk (i.e., $p_2 - p_1$) that can be obtained as follows:
\begin{Sinput}
R> plot_posterior(ab, what = "arisk")
\end{Sinput}
The top panel of Figure~\ref{fig:posteriors} shows the prior distribution as a dotted line and the posterior distribution (with 95\% central credible interval) as a solid line. The plot indicates that, under the assumption that the difference between the two success probabilities is not exactly zero, it is likely to be smaller than expected: the posterior median is $0.067$ and the 95\% central credible interval ranges from $0.011$ to $0.122$.

The middle panel of Figure~\ref{fig:posteriors} displays the posterior distribution for the log odds ratio $\psi$ that can be obtained as follows:
\begin{Sinput}
R> plot_posterior(ab, what = "logor")
\end{Sinput}
The middle panel of Figure~\ref{fig:posteriors} indicates that, given the log odds ratio is not exactly zero, it is likely to be between $0.043$ and $0.492$, where the posterior median is $0.267$.

It may also be of interest to consider the marginal posterior distributions of the success probabilities $p_1$ and $p_2$. This plot can be produced as follows:
\begin{Sinput}
R> plot_posterior(ab, what = "p1p2")
\end{Sinput}
The bottom panel of Figure~\ref{fig:posteriors} displays the resulting plot. In this example, $p_1$ and $p_2$ correspond to the probability of still being on the job after six month for the non-trained employees and the employees that received the training, respectively. The bottom panel of Figure~\ref{fig:posteriors} indicates that the posterior median for $p_1$ is $0.485$, with 95\% credible ranging from $0.443$ to $0.527$, and the posterior median for $p_2$ is $0.551$, with 95\% credible interval ranging from $0.509$ to $0.592$.

In sum, this fictitious data set offers modest evidence in favor of the null hypothesis which states that the training is not effective over the hypothesis that the training is highly effective; nevertheless, the consultancy firm should probably continue to collect data in order to obtain more compelling evidence before deciding whether or not the training should be implemented. If the true effect is as small as 4\%, continued testing will ultimately show compelling evidence for $\mathcal{H}_+$ over $\mathcal{H}_0$. Note that continued testing is trivial in the Bayesian framework: the results can simply be updated as new observations arrive.  

\section{Example 2: progesterone in women with bleeding in early pregnancy}

\begin{figure}[!tb]
	\centering
	\includegraphics[width = 0.95\textwidth]{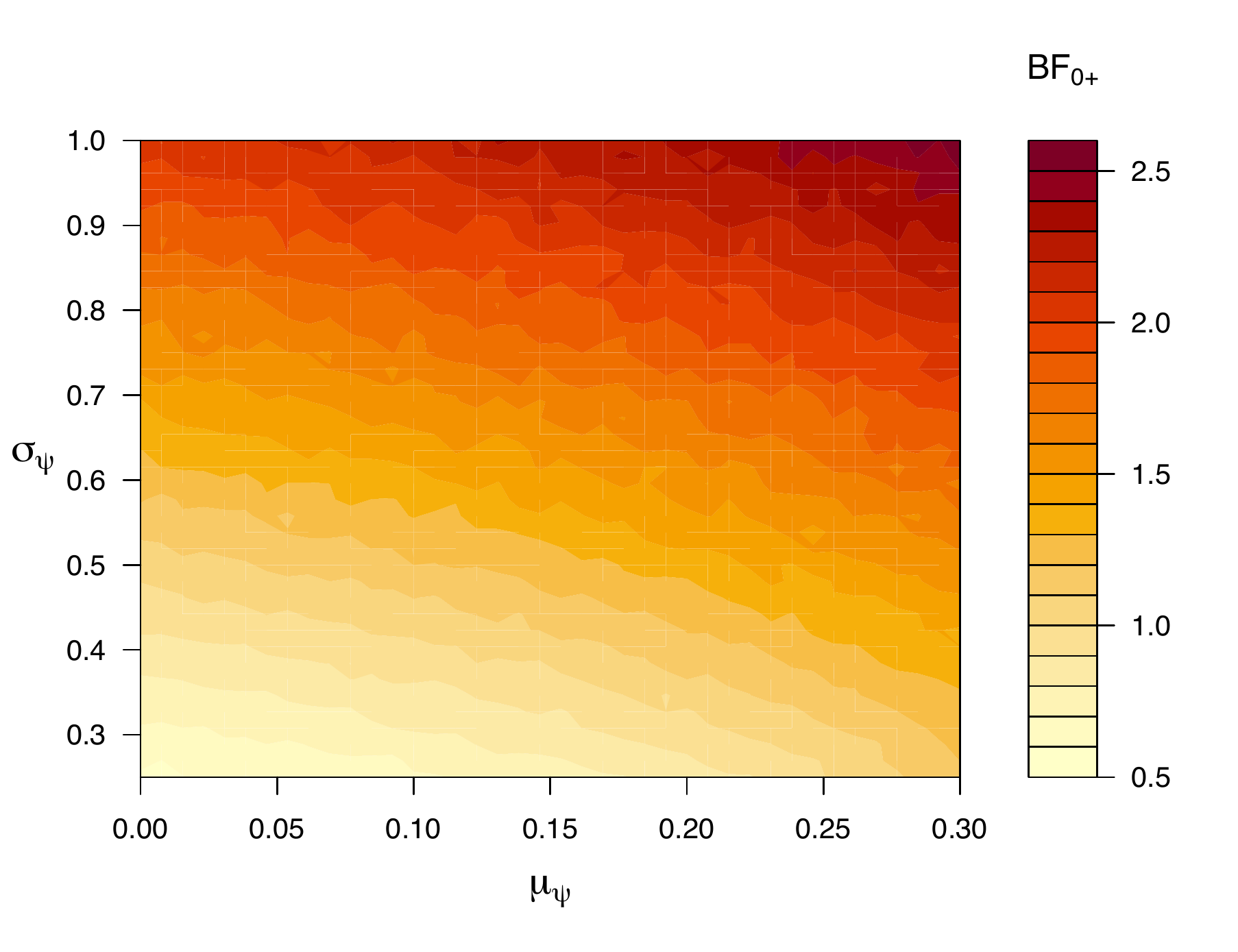}
	\caption{Prior robustness analysis for Example~2. The heat map displays the Bayes factor $\text{BF}_{0+}$ as a function of the test-relevant prior parameters $\mu_\psi$ and $\sigma_\psi$. Across different prior settings, the evidence for the no-effect hypothesis $\mathcal{H}_0$ over the positive-effect hypothesis $\mathcal{H}_+$ is weak.}
	\label{fig:robustness_progesterone}
\end{figure}

As a second example application of the \pkg{abtest} package, here we present a reanalysis of a recent medical trial.\footnote{This reanalysis is also available on \textit{PsyArXiv}: Gronau, Q. F., \& Wagenmakers, E.--J. (2019). Progesterone in women with bleeding in early pregnancy: Absence of evidence, not evidence of absence. \url{https://psyarxiv.com/etk7g/}}
\citet{CoomarasamyEtAl2019} assessed the effectiveness of progesterone in preventing miscarriages. The number of live births was 74.7\% (1513/2025) in the progesterone group and 72.5\% (1459/2013) in the placebo group ($p=.08$). The authors concluded: ``The incidence of adverse events did not differ significantly between the groups'' \citep[p. 1815]{CoomarasamyEtAl2019}.

This conclusion leaves unaddressed the degree to which the data undercut or support the no-effect hypothesis $\mathcal{H}_0$ over the positive-effect hypothesis $\mathcal{H}_+$. To quantify such evidence we can use the \pkg{abtest} package.
A default analysis can be conducted as follows:
\begin{Sinput}
R> data <- list(y1 = 1459, n1 = 2013, y2 = 1513, n2 = 2025)
R> set.seed(1)
R> ab <- ab_test(data = data)
\end{Sinput}
This yields the following output:
\begin{Soutput}
R> print(ab)

Bayesian A/B Test Results:

 Bayes Factors:

 BF10: 0.259709
 BF+0: 0.4866008
 BF-0: 0.02796485

 Prior Probabilities Hypotheses:

 H+: 0.25
 H-: 0.25
 H0: 0.5

 Posterior Probabilities Hypotheses:

 H+: 0.1935
 H-: 0.0111
 H0: 0.7954
\end{Soutput}

A Bayes factor of $\text{BF}_{0+} = 1/\text{BF}_{+0} \approx 2$ indicates that there is only weak evidence in favor of the no-effect hypothesis $\mathcal{H}_0$ over the positive-effect hypothesis $\mathcal{H}_+$ \citep{Jeffreys1939}.
To alleviate concerns about the choice of the prior distribution for the test-relevant log odds ratio parameter $\psi$ one can conduct a prior robustness analysis as follows:
\begin{Sinput}
R> plot_robustness(ab, bftype = "BF0+")
\end{Sinput}
Note that the \code{bftype} argument is used to indicate which Bayes factor is plotted (in this case $\text{BF}_{0+}$). Figure~\ref{fig:robustness_progesterone} displays the results and shows that the evidence is weak for all combinations of $\mu_\psi \in [0,0.30]$ and $\sigma_\psi \in [0.25,1]$.

In sum, these data neither undercut nor support the progesterone hypothesis in compelling fashion.

\section{Concluding comments}
In this article, we have introduced the \pkg{abtest} package that implements both Bayesian hypothesis testing and Bayesian estimation for the A/B test using informed priors. The procedure allows users to (1) obtain evidence in favor of the null hypothesis; (2) monitor the evidence as data accumulate; and (3) elicit and incorporate expert prior distributions. We hope that the provided analysis approach is useful across different fields that apply A/B testing on a routine basis, particularly business and medicine.

\begin{figure}[!tb]
	\centering
	\includegraphics[width = \textwidth]{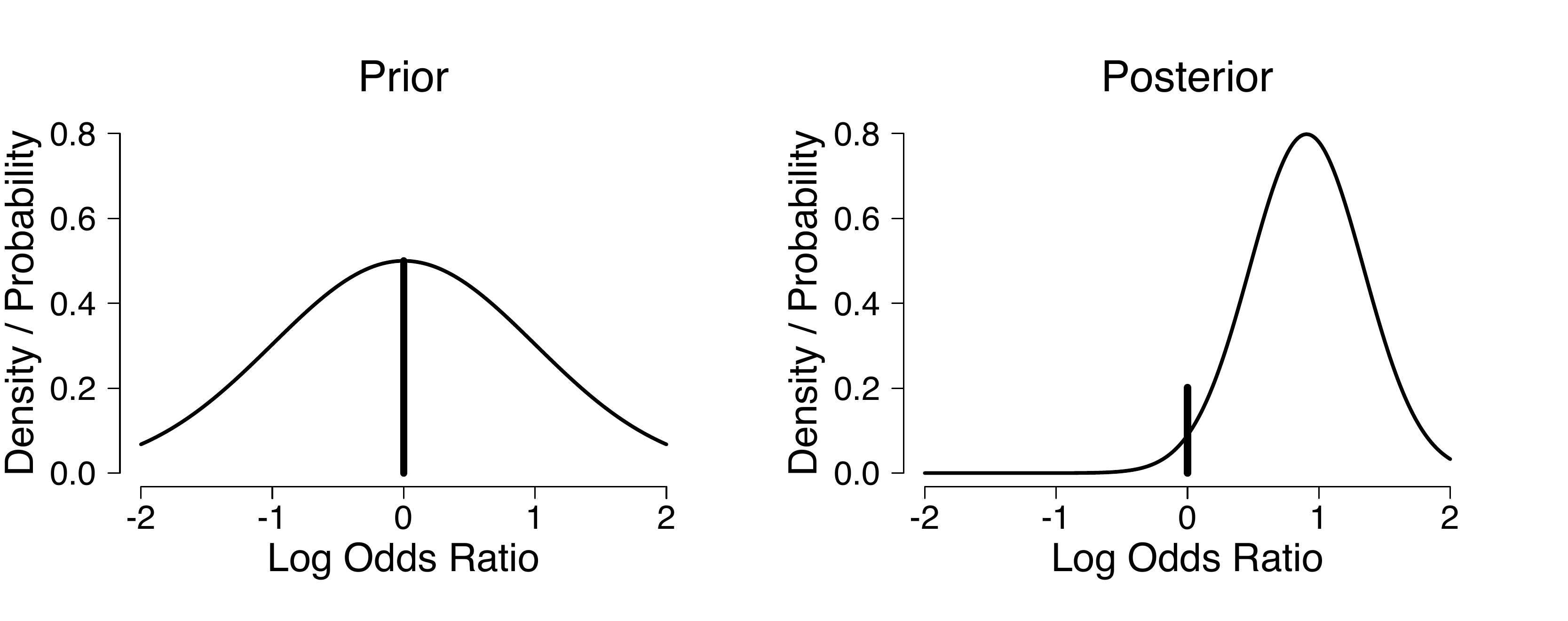}
	\caption{Mixture representation of the A/B test procedure. The left panel illustrates the mixture representation before having observed any data, the right panel illustrates the mixture representation after having observed 20 successes out of 40 observations in the control condition and 30 successes out of 40 observations in the experimental condition. The height of the spike at zero corresponds to the prior/posterior probability of $\mathcal{H}_0$ whereas the shape of the slab corresponds to the continuous default prior/posterior distribution for $\psi$ under $\mathcal{H}_1$. The maximum height of this continuous distribution corresponds to the prior/posterior probability of $\mathcal{H}_1$.}
	\label{fig:mixture_plot}
\end{figure}

We have introduced the approach implemented in \pkg{abtest} as testing hypotheses of interest about the test-relevant log odds ratio parameter $\psi$ for the model in Equation~\ref{eq:model}. However, it should be pointed out that an alternative interpretation is to view the procedure as estimating a mixture model, where the mixture components correspond to the different hypotheses of interest, and the mixture weights are given by the prior/posterior probabilities of the hypotheses \citep[e.g.,][]{MitchellBeauchamp1988}. This interpretation is illustrated with a fictitious example in Figure~\ref{fig:mixture_plot}. For simplicity, the plot assumes that the user has set the prior probabilities of $\mathcal{H}_+$ and $\mathcal{H}_-$  to zero, whereas the prior probabilities of $\mathcal{H}_1$ and $\mathcal{H}_0$ are both set to .50. The left panel illustrates the mixture representation before having observed any data. Specifically, the height of the spike at zero corresponds to the prior probability of $\mathcal{H}_0$ whereas the shape of the slab corresponds to the continuous default prior distribution for $\psi$ under $\mathcal{H}_1$. The maximum height of this continuous distribution corresponds to the prior probability of $\mathcal{H}_1$.\footnote{This scaling method is inspired by the \pkg{BAS} package \citep{BAS}.} The right panel illustrates the mixture representation after having observed 20 successes out of 40 observations in the control condition and 30 successes out of 40 observations in the experimental condition (these are fictitious data). The height of the spike corresponds to the posterior probability of $\mathcal{H}_0$, and the maximum height of the continuous posterior distribution under $\mathcal{H}_1$ (i.e., the slab) corresponds to the posterior probability of $\mathcal{H}_1$. In this fictitious example, the data have decreased the plausibility of $\mathcal{H}_0$ and have increased the plausibility of $\mathcal{H}_1$.

Despite the practical benefits that the package offers right now, there are areas for future improvement. For instance, \pkg{abtest} currently allows users to compare two groups; however, there are applications in which one may be interested in simultaneously comparing more than two groups. Furthermore, at the moment, \pkg{abtest} expects the outcome variable to be binary. Nevertheless, in certain scenarios, it may be more natural to compare the two groups based on a continuous outcome variable. This scenario resembles an independent samples $t$-test for which well-established Bayesian procedures exist \citep[e.g.,][]{RouderEtAl2009Ttest,LyEtAl2016} which are available, for instance, in the \pkg{BayesFactor} package \citep{BayesFactor} and \proglang{JASP} \citep{JASP}.\footnote{For a list of Bayesian \proglang{R} packages, see \url{https://cran.r-project.org/web/views/Bayesian.html}.} Moreover, currently, the \pkg{abtest} package does not provide functions for generating predictions. Note, however, that users can generate predictions in a straightforward manner themselves based on the posterior samples that are provided by \pkg{abtest}. The implementation also does not allow users to incorporate utilities explicitly (e.g., \citealp{Lindley1985}; for alternative approaches see also \citealp{AzevedoEtAl2019} and \citealp{FeitBerman2019}). However, again, based on the provided posterior probabilities and posterior samples, users who wish to take into account utilities may do so in a relatively straightforward way.
Furthermore, users interested in adjusting the model used in \pkg{abtest} (e.g., to account for hierarchically-structured data or covariates) are referred to general-purpose Bayesian software such as \proglang{Stan} \citep{CarpenterEtAl2017, rstan} and the related \proglang{R} package \pkg{brms} \citep{brms}. In combination with the \pkg{bridgesampling} package \citep{GronauEtAlbridgesampling}, this enables the user to compare custom models using Bayes factors and posterior model probabilities. A more structural limitation of \pkg{abtest} is that it has been developed to analyze A/B test data, but not to run the A/B test experiment itself.

In sum, A/B testing is ubiquitous in business and medicine. Here we have demonstrated how the \pkg{abtest} package enables relatively complete Bayesian inference including the capability to obtain support for the null, continuously monitor the results, and elicit and incorporate expert prior knowledge. Hopefully, this approach forms a basis for evidence-based conclusions that will benefit both businesses and patients.

\section{Acknowledgements}
This research was supported by a Netherlands Organisation for Scientific Research (NWO) grant to QFG (406.16.528) and by an NWO Vici grant to EJW (016.Vici.170.083). 



\bibliography{references}

\begin{thebibliography}{52}
\newcommand{\enquote}[1]{``#1''}
\providecommand{\natexlab}[1]{#1}
\providecommand{\url}[1]{\texttt{#1}}
\providecommand{\urlprefix}{URL }
\expandafter\ifx\csname urlstyle\endcsname\relax
  \providecommand{\doi}[1]{doi:\discretionary{}{}{}#1}\else
  \providecommand{\doi}{doi:\discretionary{}{}{}\begingroup
  \urlstyle{rm}\Url}\fi
\providecommand{\eprint}[2][]{\url{#2}}

\bibitem[{Armitage(1960)}]{Armitage1960}
Armitage P (1960).
\newblock \emph{Sequential Medical Trials}.
\newblock Thomas, Springfield (IL).

\bibitem[{Azevedo \emph{et~al.}(2019)Azevedo, Alex, Montiel~Olea, Rao, and
  Weyl}]{AzevedoEtAl2019}
Azevedo EM, Alex D, Montiel~Olea J, Rao JM, Weyl EG (2019).
\newblock \enquote{{A/B} Testing with Fat Tails.}
\newblock \emph{SSRN}.
\newblock \urlprefix\url{http://dx.doi.org/10.2139/ssrn.3171224}.

\bibitem[{B{\aa}{\aa}th(2014)}]{Baath2014BayesianFirstAid}
B{\aa}{\aa}th R (2014).
\newblock \enquote{Bayesian First Aid: A Package that Implements Bayesian
  Alternatives to the Classical \code{*.test} Functions in \proglang{R}.}
\newblock In \emph{UseR! 2014 - the International \proglang{R} User
  Conference}.

\bibitem[{Bartlett(1957)}]{Bartlett1957}
Bartlett MS (1957).
\newblock \enquote{A Comment on {D. V. Lindley}'s Statistical Paradox.}
\newblock \emph{Biometrika}, \textbf{44}, 533--534.

\bibitem[{Berger and Delampady(1987)}]{BergerDelampady1987}
Berger JO, Delampady M (1987).
\newblock \enquote{Testing Precise Hypotheses.}
\newblock \emph{Statistical Science}, \textbf{2}, 317--352.

\bibitem[{Berger and Wolpert(1988)}]{BergerWolpert1988}
Berger JO, Wolpert RL (1988).
\newblock \emph{The Likelihood Principle (2nd ed.)}.
\newblock Institute of Mathematical Statistics, Hayward (CA).

\bibitem[{Berman \emph{et~al.}(2018)Berman, Pekelis, Scott, and Van~den
  Bulte}]{BermanEtAl2018}
Berman R, Pekelis L, Scott A, Van~den Bulte C (2018).
\newblock \enquote{p-Hacking and False Discovery in {A/B} Testing.}
\newblock \emph{SSRN}.
\newblock \urlprefix\url{http://dx.doi.org/10.2139/ssrn.3204791}.

\bibitem[{Bürkner(2017)}]{brms}
Bürkner PC (2017).
\newblock \enquote{\pkg{brms}: An \proglang{R} Package for {Bayesian}
  Multilevel Models Using \proglang{Stan}.}
\newblock \emph{Journal of Statistical Software}, \textbf{80}, 1--28.

\bibitem[{Carpenter \emph{et~al.}(2017)Carpenter, Gelman, Hoffman, Lee,
  Goodrich, Betancourt, Brubaker, Guo, Li, and Riddell}]{CarpenterEtAl2017}
Carpenter B, Gelman A, Hoffman M, Lee D, Goodrich B, Betancourt M, Brubaker M,
  Guo J, Li P, Riddell A (2017).
\newblock \enquote{\proglang{Stan}: A Probabilistic Programming Language.}
\newblock \emph{Journal of Statistical Software}, \textbf{76}, 1--32.

\bibitem[{Chen \emph{et~al.}(2010)Chen, Cohen, and Chen}]{ChenEtAl2010}
Chen H, Cohen P, Chen S (2010).
\newblock \enquote{How Big Is a Big Odds Ratio? Interpreting the Magnitudes of
  Odds Ratios in Epidemiological Studies.}
\newblock \emph{Communications in Statistics---Simulation and
  Computation{\textregistered}}, \textbf{39}, 860--864.

\bibitem[{Clyde(2020)}]{BAS}
Clyde M (2020).
\newblock \emph{\pkg{BAS}: Bayesian Variable Selection and Model Averaging
  Using Bayesian Adaptive Sampling}.
\newblock \proglang{R} package version 1.5.5.

\bibitem[{Coomarasamy \emph{et~al.}(2019)Coomarasamy, Devall, Cheed, Harb,
  Middleton, Gallos, Williams, Eapen, Roberts, Ogwulu, Goranitis, Daniels,
  Ahmed, Bender‑Atik, Bhatia, Bottomley, Brewin, Choudhary, Crosfill, Deb,
  Duncan, Ewer, Hinshaw, Holland, Izzat, Johns, Kriedt, Lumsden, Manda, Norman,
  Nunes, Overton, Quenby, Rao, Ross, Shahid, Underwood, Vaithilingam, Watkins,
  Wykes, Horne, and Jurkovic}]{CoomarasamyEtAl2019}
Coomarasamy A, Devall AJ, Cheed V, Harb H, Middleton LJ, Gallos ID, Williams H,
  Eapen AK, Roberts T, Ogwulu CC, Goranitis I, Daniels JP, Ahmed A,
  Bender‑Atik R, Bhatia K, Bottomley C, Brewin J, Choudhary M, Crosfill F,
  Deb S, Duncan WC, Ewer A, Hinshaw K, Holland T, Izzat F, Johns J, Kriedt K,
  Lumsden MA, Manda P, Norman JE, Nunes N, Overton CE, Quenby S, Rao S, Ross J,
  Shahid A, Underwood M, Vaithilingam N, Watkins L, Wykes C, Horne A, Jurkovic
  D (2019).
\newblock \enquote{A Randomized Trial of Progesterone in Women with Bleeding in
  Early Pregnancy.}
\newblock \emph{New England Journal of Medicine}, \textbf{380}, 1815--1824.

\bibitem[{Cumming(2014)}]{Cumming2014}
Cumming G (2014).
\newblock \enquote{The New Statistics: {W}hy and How.}
\newblock \emph{Psychological Science}, \textbf{25}, 7--29.

\bibitem[{Deng \emph{et~al.}(2016)Deng, Lu, and Chen}]{DengEtAl2016}
Deng A, Lu J, Chen S (2016).
\newblock \enquote{Continuous Monitoring of {A/B} Tests Without Pain: Optional
  Stopping in Bayesian Testing.}
\newblock In \emph{2016 IEEE International Conference on Data Science and
  Advanced Analytics}, pp. 243--252.

\bibitem[{Dienes(2014)}]{Dienes2014}
Dienes Z (2014).
\newblock \enquote{Using {B}ayes to Get the Most out of Non-significant
  Results.}
\newblock \emph{Frontiers in Psycholology}, \textbf{{5:781}}.

\bibitem[{Feit and Berman(2019)}]{FeitBerman2019}
Feit EM, Berman R (2019).
\newblock \enquote{Test \& Roll: {P}rofit-Maximizing {A/B} Tests.}
\newblock \emph{Marketing Science}, \textbf{38}, 1038--1058.

\bibitem[{Feller(1940)}]{Feller1940}
Feller W (1940).
\newblock \enquote{Statistical Aspects of {ESP}.}
\newblock \emph{Journal of Parapsychology}, \textbf{4}, 271--298.

\bibitem[{Fisher(1928)}]{Fisher1928}
Fisher RA (1928).
\newblock \emph{Statistical Methods for Research Workers}.
\newblock 2nd edition. Oliver and Boyd, Edinburgh.

\bibitem[{Gelman and Rubin(1995)}]{GelmanRubin1995}
Gelman A, Rubin DB (1995).
\newblock \enquote{Avoiding Model Selection in {B}ayesian Social Research.}
\newblock \emph{Sociological Methodology}, \textbf{25}, 165--173.

\bibitem[{Gronau \emph{et~al.}(2020)Gronau, Singmann, and
  Wagenmakers}]{GronauEtAlbridgesampling}
Gronau QF, Singmann H, Wagenmakers EJ (2020).
\newblock \enquote{\pkg{bridgesampling}: {A}n \proglang{R} Package for
  Estimating Normalizing Constants.}
\newblock \emph{Journal of Statistical Software}, \textbf{92}.
\newblock \urlprefix\url{https://www.jstatsoft.org/article/view/v092i10}.

\bibitem[{Haaf \emph{et~al.}(2019)Haaf, Ly, and Wagenmakers}]{HaafEtAl2019}
Haaf J, Ly A, Wagenmakers EJ (2019).
\newblock \enquote{Retire Significance, but Still Test Hypotheses.}
\newblock \emph{Nature}, \textbf{567}, 461.

\bibitem[{Howard(1998)}]{Howard1998}
Howard JV (1998).
\newblock \enquote{The $2\times2$ Table: {A} Discussion from a {B}ayesian
  Viewpoint.}
\newblock \emph{Statistical Science}, \textbf{13}, 351--367.

\bibitem[{Jamil \emph{et~al.}(2017)Jamil, Marsman, Ly, Morey, and
  Wagenmakers}]{JamilEtAl2017}
Jamil T, Marsman M, Ly A, Morey RD, Wagenmakers EJ (2017).
\newblock \enquote{What Are the Odds? {M}odern Relevance and {B}ayes Factor
  Solutions for {MacAlister}'s Problem from the 1881 \emph{Educational Times}.}
\newblock \emph{Educational and Psychological Measurement}, \textbf{77},
  819--830.

\bibitem[{Jeffreys(1935)}]{Jeffreys1935}
Jeffreys H (1935).
\newblock \enquote{Some Tests of Significance, Treated by the Theory of
  Probability.}
\newblock \emph{Proceedings of the Cambridge Philosophy Society}, \textbf{31},
  203--222.

\bibitem[{Jeffreys(1939)}]{Jeffreys1939}
Jeffreys H (1939).
\newblock \emph{Theory of Probability}.
\newblock 1st edition. Oxford University Press, Oxford, UK.

\bibitem[{Johari \emph{et~al.}(2017)Johari, Koomen, Pekelis, and
  Walsh}]{JohariEtAl2017}
Johari R, Koomen P, Pekelis L, Walsh D (2017).
\newblock \enquote{Peeking at A/B Tests: Why It Matters, and What to Do About
  It.}
\newblock In \emph{Proceedings of the 23rd ACM SIGKDD International Conference
  on Knowledge Discovery and Data Mining}, KDD '17, pp. 1517--1525. ACM, New
  York, NY, USA.
\newblock \urlprefix\url{http://doi.acm.org/10.1145/3097983.3097992}.

\bibitem[{Kass and Raftery(1995)}]{KassRaftery1995}
Kass RE, Raftery AE (1995).
\newblock \enquote{{B}ayes Factors.}
\newblock \emph{Journal of the American Statistical Association}, \textbf{90},
  773--795.

\bibitem[{Kass and Vaidyanathan(1992)}]{KassVaidyanathan1992}
Kass RE, Vaidyanathan SK (1992).
\newblock \enquote{Approximate {B}ayes Factors and Orthogonal Parameters, with
  Application to Testing Equality of Two Binomial Proportions.}
\newblock \emph{Journal of the Royal Statistical Society, Series B},
  \textbf{54}, 129--144.

\bibitem[{Keysers \emph{et~al.}(2020)Keysers, Gazzola, and
  Wagenmakers}]{KeysersEtAl2020}
Keysers C, Gazzola V, Wagenmakers EJ (2020).
\newblock \enquote{Using {B}ayes Factor Hypothesis Testing in Neuroscience to
  Establish Evidence of Absence.}
\newblock \emph{Nature Neuroscience}, \textbf{23}, 788--799.

\bibitem[{Lindley(1957)}]{Lindley1957}
Lindley DV (1957).
\newblock \enquote{A Statistical Paradox.}
\newblock \emph{Biometrika}, \textbf{44}, 187--192.

\bibitem[{Lindley(1985)}]{Lindley1985}
Lindley DV (1985).
\newblock \emph{Making Decisions}.
\newblock 2nd edition. John Wiley \& Sons, London.

\bibitem[{Lipkus and Hollands(1999)}]{LipkusHollands1999}
Lipkus IM, Hollands JG (1999).
\newblock \enquote{The Visual Communication of Risk.}
\newblock \emph{Journal of the National Cancer Institute Monographs},
  \textbf{25}, 149--163.

\bibitem[{Little(1989)}]{Little1989}
Little RJA (1989).
\newblock \enquote{Testing the Equality of Two Independent Binomial
  Proportions.}
\newblock \emph{The American Statistician}, \textbf{43}, 283--288.

\bibitem[{Ly \emph{et~al.}(2016)Ly, Verhagen, and Wagenmakers}]{LyEtAl2016}
Ly A, Verhagen AJ, Wagenmakers EJ (2016).
\newblock \enquote{Harold {J}effreys's Default {B}ayes Factor Hypothesis Tests:
  {E}xplanation, Extension, and Application in Psychology.}
\newblock \emph{Journal of Mathematical Psychology}, \textbf{72}, 19--32.

\bibitem[{Malek \emph{et~al.}(2017)Malek, Katariya, Chow, and
  Ghavamzadeh}]{MalekEtAl2017}
Malek A, Katariya S, Chow Y, Ghavamzadeh M (2017).
\newblock \enquote{Sequential Multiple Hypothesis Testing with Type I Error
  Control.}
\newblock In \emph{Proceedings of the 20th International Conference on
  Artificial Intelligence and Statistics}, pp. 1468--1476.

\bibitem[{Meng and Wong(1996)}]{MengWong1996}
Meng XL, Wong WH (1996).
\newblock \enquote{Simulating Ratios of Normalizing Constants via a Simple
  Identity: {A} Theoretical Exploration.}
\newblock \emph{Statistica Sinica}, \textbf{6}, 831--860.

\bibitem[{Mitchell and Beauchamp(1988)}]{MitchellBeauchamp1988}
Mitchell TJ, Beauchamp JJ (1988).
\newblock \enquote{{B}ayesian Variable Selection in Linear Regression.}
\newblock \emph{Journal of the American Statistical Association}, \textbf{83},
  1023--1032.

\bibitem[{Morey and Rouder(2018)}]{BayesFactor}
Morey RD, Rouder JN (2018).
\newblock \emph{\pkg{BayesFactor}: Computation of {B}ayes Factors for Common
  Designs}.
\newblock \proglang{R} package version 0.9.12-4.2,
  \urlprefix\url{https://CRAN.R-project.org/package=BayesFactor}.

\bibitem[{O'Hagan(2019)}]{OHagan20019}
O'Hagan A (2019).
\newblock \enquote{Expert Knowledge Elicitation: Subjective but Scientific.}
\newblock \emph{The American Statistician}, \textbf{73}, 69--81.

\bibitem[{Pham-Gia \emph{et~al.}(2017)Pham-Gia, Van~Thin, and
  Doan}]{PhamEtAl2017}
Pham-Gia T, Van~Thin N, Doan PP (2017).
\newblock \enquote{Inferences on the Difference of Two Proportions: A Bayesian
  Approach.}
\newblock \emph{Open Journal of Statistics}, \textbf{7}, 1--15.

\bibitem[{Portman(2019)}]{bayesAB}
Portman F (2019).
\newblock \emph{\pkg{bayesAB}: Fast {B}ayesian Methods for {AB} Testing}.
\newblock \proglang{R} package version 1.1.2,
  \urlprefix\url{https://CRAN.R-project.org/package=bayesAB}.

\bibitem[{\proglang{JASP} Team(2020)}]{JASP}
\proglang{JASP} Team (2020).
\newblock \enquote{\proglang{JASP} (Version 0.14)[Computer software].}
\newblock \urlprefix\url{https://jasp-stats.org/}.

\bibitem[{\proglang{R} Core~Team(2019)}]{R}
\proglang{R} Core~Team (2019).
\newblock \emph{\proglang{R}: A Language and Environment for Statistical
  Computing}.
\newblock \proglang{R} Foundation for Statistical Computing, Vienna, Austria.
\newblock \urlprefix\url{https://www.R-project.org/}.

\bibitem[{\proglang{Stan}~{D}evelopment {T}eam(2019)}]{rstan}
\proglang{Stan}~{D}evelopment {T}eam (2019).
\newblock \enquote{\pkg{rstan}: the \proglang{R} interface to \proglang{Stan}.}
\newblock \proglang{R} package version 2.19.2,
  \urlprefix\url{http://mc-stan.org/}.

\bibitem[{Robert and Casella(2010)}]{RobertCasella2010}
Robert C, Casella G (2010).
\newblock \emph{Introducing {M}onte {C}arlo Methods with \proglang{R}}.
\newblock Springer-Verlag, New York.

\bibitem[{Rouder(2014)}]{Rouder2014PBR}
Rouder JN (2014).
\newblock \enquote{Optional Stopping: {N}o Problem for {B}ayesians.}
\newblock \emph{Psychonomic Bulletin \& Review}, \textbf{21}, 301--308.

\bibitem[{Rouder \emph{et~al.}(2009)Rouder, Speckman, Sun, Morey, and
  Iverson}]{RouderEtAl2009Ttest}
Rouder JN, Speckman PL, Sun D, Morey RD, Iverson G (2009).
\newblock \enquote{{B}ayesian $T$ Tests for Accepting and Rejecting the Null
  Hypothesis.}
\newblock \emph{Psychonomic Bulletin \& Review}, \textbf{16}, 225--237.

\bibitem[{Skorski(2019)}]{Skorski2019}
Skorski M (2019).
\newblock \enquote{Bounds on Bayes Factors for Binomial {A/B} Testing.}
\newblock \emph{arXiv preprint arXiv:1903.00049}.
\newblock \urlprefix\url{https://arxiv.org/abs/1903.00049}.

\bibitem[{Stucchio(2015)}]{Stucchio2015}
Stucchio C (2015).
\newblock \enquote{Bayesian {A/B} Testing at {VWO}.}
\newblock \emph{Technical report}, VWO.
\newblock
  \urlprefix\url{https://www.chrisstucchio.com/pubs/VWO_SmartStats_technical_whitepaper.pdf}.

\bibitem[{Tversky(1969)}]{Tversky1969}
Tversky A (1969).
\newblock \enquote{Intransitivity of Preferences.}
\newblock \emph{Psychological Review}, \textbf{76}, 31--48.

\bibitem[{Wagenmakers \emph{et~al.}(2018)Wagenmakers, Marsman, Jamil, Ly,
  Verhagen, Love, Selker, Gronau, \v{S}m\'{i}ra, Epskamp, Matzke, Rouder, and
  Morey}]{WagenmakersEtAl2018PBRPartI}
Wagenmakers EJ, Marsman M, Jamil T, Ly A, Verhagen AJ, Love J, Selker R, Gronau
  QF, \v{S}m\'{i}ra M, Epskamp S, Matzke D, Rouder JN, Morey RD (2018).
\newblock \enquote{Bayesian Inference for Psychology. {Part I}: {T}heoretical
  Advantages and Practical Ramifications.}
\newblock \emph{Psychonomic Bulletin \& Review}, \textbf{25}, 35--57.

\bibitem[{Ware(1989)}]{Ware1989}
Ware JH (1989).
\newblock \enquote{Investigating Therapies of Potentially Great Benefit:
  {ECMO}.}
\newblock \emph{Statistical Science}, \textbf{4}, 298--340.

\end{thebibliography}

\appendix

\section{Interpretation of the parameters}
Here we show that $\beta$ corresponds to the grand mean of the log odds and that $\psi$ corresponds to the log odds ratio (for the model definition, see Equation~\ref{eq:model}).
The nuisance parameter $\beta$ corresponds to the grand mean of the log odds since
\begin{equation*}
\frac{1}{2} \log\left(\frac{p_1}{1 - p_1}\right) + \frac{1}{2} \log\left(\frac{p_2}{1 - p_2}\right) = \frac{1}{2} \beta - \frac{1}{4} \psi + \frac{1}{2} \beta + \frac{1}{4} \psi = \beta.
\end{equation*}
The test-relevant parameter $\psi$ corresponds to the log odds ratio since
\begin{equation*}
\log\left(\frac{\frac{p_2}{1 - p_2}}{\frac{p_1}{1 - p_1}}\right) = \log\left(\frac{p_2}{1 - p_2}\right) - \log\left(\frac{p_1}{1 - p_1}\right) = \beta + \frac{\psi}{2} - \left(\beta - \frac{\psi}{2}\right) = \psi.
\end{equation*}

\section{Prior elicitation: implied distributions}
The prior elicitation approach described in Equation~\ref{eq:elicitation} requires the cdf's for the quantities of interest. Here, we derive the implied cdf's for these quantities; we also derive the corresponding probability density functions (pdf's). Additionally, we derive four further implied distributions of interest: the joint pdf of $p_1$ and $p_2$, the conditional pdf of $p_2$ given $p_1$ is fixed to a particular value, the marginal distribution for $p_1$, and the marginal distribution for $p_2$. A few of these expressions will contain a one-dimensional integral which can easily be evaluated using numerical integration.

\subsection{Log odds ratio}
Since $\psi$ itself corresponds to the log odds ratio, $F(\cdot; \mu_\psi, \sigma_\psi)$ corresponds in this case to the cdf of a normal distribution with mean $\mu_\psi$ and standard deviation $\sigma_\psi$. The corresponding pdf is the normal probability density function.

\subsection{Odds ratio}
The implied prior on the odds ratio $\omega = \exp(\psi)$ is a log-normal distribution. Hence, $F(\cdot; \mu_\psi, \sigma_\psi)$ corresponds in this case to the cdf of a log-normal distribution with parameters $\mu_\psi$ and $\sigma_\psi$. The corresponding pdf is the log-normal probability density function.

\subsection{Relative risk}
The relative risk is given by $\Lambda = \frac{p_2}{p_1}$.
We use a capital letter (i.e., $\Lambda$) to refer to the random variable and use a lower-case letter (i.e., $\lambda$) to refer to a concrete realization.
Note that so far, we have abused notation by only using lower-case letters, but it should be clear from the context when we referred to a random variable or a concrete realization. However, for deriving the following cdf, we need the distinction to keep the notation clear.
To derive the implied cdf for the relative risk, we proceed as follows:
\begin{align*}
P(\Lambda \le \lambda) &= P\left(\frac{p_2}{p_1} \le \lambda\right) \\
&= P\left(p_2 \le \lambda p_1 \right) \\
&= P\left(\frac{1}{1 + \exp\left(-\beta - \frac{\psi}{2}\right)} \le  \frac{\lambda}{1 + \exp\left(-\beta + \frac{\psi}{2}\right)}\right).
\end{align*}
Taking reciprocals and some algebra yields
\begin{align*}
P\left(\left(\exp\left(\frac{\psi}{2}\right)\right)^2 + \left(1 - \lambda\right) \exp(\beta)\exp\left(\frac{\psi}{2}\right) - \lambda \le  0\right).
\end{align*}
When we set
\begin{align*}
\left(\exp\left(\frac{\psi}{2}\right)\right)^2 + \left(1 - \lambda\right) \exp(\beta)\exp\left(\frac{\psi}{2}\right) - \lambda =  0,
\end{align*}
we can solve for $\psi$ using the fact that this is a quadratic equation in $\exp\left(\frac{\psi}{2}\right)$ and we obtain:
\begin{align*}
\exp\left(\frac{\psi}{2}\right) = \frac{-\left(1 - \lambda\right)\exp(\beta) + \sqrt{\left(1 - \lambda\right)^2\exp(2\beta) + 4\lambda}}{2},
\end{align*}
where we took into account that $\exp\left(\frac{\psi}{2}\right)$ needs to be positive (i.e., we omitted the solution corresponding to minus the square root).
Hence,
\begin{align*}
\psi = 2 \log\left(\frac{-\left(1 - \lambda\right)\exp(\beta) + \sqrt{\left(1 - \lambda\right)^2\exp(2\beta) + 4\lambda}}{2}\right).
\end{align*}
Therefore, $\left(\exp\left(\frac{\psi}{2}\right)\right)^2 + \left(1 - \lambda\right) \exp(\beta)\exp\left(\frac{\psi}{2}\right) - \lambda \le  0$ whenever 
\begin{equation*}
\psi \le 2 \log\left(\frac{-\left(1 - \lambda\right)\exp(\beta) + \sqrt{\left(1 - \lambda\right)^2\exp(2\beta) + 4\lambda}}{2}\right).
\end{equation*}
Hence, the desired cdf can be written as
\begin{equation}
\begin{split}
&P\left(\psi \le 2 \log\left(\frac{-\left(1 - \lambda\right)\exp(\beta) + \sqrt{\left(1 - \lambda\right)^2\exp(2\beta) + 4\lambda}}{2}\right)\right) \\
=& \int_{-\infty}^{\infty} \int_{-\infty}^{2 \log\left(\frac{-\left(1 - \lambda\right)\exp(\beta) + \sqrt{\left(1 - \lambda\right)^2\exp(2\beta) + 4\lambda}}{2}\right)} \mathcal{N}(\psi; \mu_\psi, \sigma_\psi^2) \mathcal{N}(\beta; \mu_\beta, \sigma_\beta^2) \text{d}\psi \text{d}\beta \\
=& \int_{-\infty}^{\infty} \mathcal{N}(\beta; \mu_\beta, \sigma_\beta^2) \, \Phi\left(2 \log\left(\frac{-\left(1 - \lambda\right)\exp(\beta) + \sqrt{\left(1 - \lambda\right)^2\exp(2\beta) + 4\lambda}}{2}\right); \mu_\psi, \sigma_\psi^2\right) \text{d}\beta,
\end{split}
\end{equation}
where $\Phi\left(\cdot; \mu_\psi, \sigma_\psi^2\right)$ denotes the cdf of a normal distribution with mean $\mu_\psi$ and variance $\sigma_\psi^2$, and $\mathcal{N}(\cdot; \mu_\beta, \sigma_\beta^2)$ denotes the corresponding pdf. 

The pdf of the relative risk is obtained by taking the derivative with respect to $\lambda$:
\begin{equation}
\begin{split}
\frac{d}{d \lambda} &\left[\int_{-\infty}^{\infty} \mathcal{N}(\beta; \mu_\beta, \sigma_\beta^2) \, \Phi\left(2 \log\left(\frac{-\left(1 - \lambda\right)\exp(\beta) + \sqrt{\left(1 - \lambda\right)^2\exp(2\beta) + 4\lambda}}{2}\right); \mu_\psi, \sigma_\psi^2\right) \text{d}\beta\right] \\
&= \int_{-\infty}^{\infty} \mathcal{N}(\beta; \mu_\beta, \sigma_\beta^2) \, \mathcal{N}\left(2 \log\left(\frac{-\left(1 - \lambda\right)\exp(\beta) + \sqrt{\left(1 - \lambda\right)^2\exp(2\beta) + 4\lambda}}{2}\right); \mu_\psi, \sigma_\psi^2\right) \\
& \hspace{2em} \times 2\Bigg[\frac{\exp(\beta) + \frac{2 - (1 - \lambda) \exp(2 \beta)}{\sqrt{(1 - \lambda)^2 \exp(2\beta) + 4 \lambda}}}{-(1 - \lambda) \exp(\beta) + \sqrt{(1 - \lambda)^2 \exp(2 \beta) + 4 \lambda}}\Bigg] \text{d}\beta.
\end{split}
\end{equation}

\subsection{Absolute risk}
The absolute risk is given by $\Upsilon = p_2 - p_1$.
We use  the upper-case letter $\Upsilon$ to refer to the random variable and the lower-case letter $\upsilon$ to refer to a concrete realization.
To derive the implied cdf for the absolute risk, we proceed as follows:
\begin{align*}
P(\Upsilon \le \upsilon) &= P\left(p_2 - p_1 \le \upsilon\right) \\
&= P\left(p_2 \le \upsilon + p_1 \right) \\
&= P\left(\frac{1}{1 + \exp\left(-\beta - \frac{\psi}{2}\right)} \le  \upsilon + \frac{1}{1 + \exp\left(-\beta + \frac{\psi}{2}\right)}\right).
\end{align*}
After some algebra, we obtain
\begin{align*}
P\left(\exp\left(\beta\right) \left(1 - \upsilon\right) \left(\exp\left(\frac{\psi}{2}\right)\right)^2 - \upsilon \left(\exp\left(2 \beta\right) + 1\right)  \exp\left(\frac{\psi}{2}\right)  - \exp\left(\beta\right) \left(\upsilon + 1\right) \le  0 \right).
\end{align*}
When we set
\begin{align*}
\exp\left(\beta\right) \left(1 - \upsilon\right) \left(\exp\left(\frac{\psi}{2}\right)\right)^2 - \upsilon \left(\exp\left(2 \beta\right) + 1\right)  \exp\left(\frac{\psi}{2}\right)  - \exp\left(\beta\right) \left(\upsilon + 1\right) =  0,
\end{align*}
we can solve for $\psi$ using the fact that this is a quadratic equation in $\exp\left(\frac{\psi}{2}\right)$ and we obtain:
\begin{align*}
\exp\left(\frac{\psi}{2}\right) &= \frac{\upsilon \left(\exp\left(2 \beta\right) + 1\right) + \sqrt{\upsilon^2 \left(\exp\left(2 \beta\right) - 1\right)^2 + 4 \exp\left(2 \beta\right)}}{2 \exp\left(\beta\right) \left(1 - \upsilon\right)} ,
\end{align*}
where we took into account that $\exp\left(\frac{\psi}{2}\right)$ needs to be positive (i.e., we omitted the solution corresponding to minus the square root).
Hence,
\begin{align*}
\psi = 2 \log\left(\frac{\upsilon \left(\exp\left(2 \beta\right) + 1\right) + \sqrt{\upsilon^2 \left(\exp\left(2 \beta\right) - 1\right)^2 + 4 \exp\left(2 \beta\right)}}{2 \exp\left(\beta\right) \left(1 - \upsilon\right)}\right).
\end{align*}
Therefore, $\exp\left(\beta\right) \left(1 - \upsilon\right) \left(\exp\left(\frac{\psi}{2}\right)\right)^2 - \upsilon \left(\exp\left(2 \beta\right) + 1\right)  \exp\left(\frac{\psi}{2}\right)  - \exp\left(\beta\right) \left(\upsilon + 1\right) \le  0$ whenever
\begin{equation*}
\psi \le 2 \log\left(\frac{\upsilon \left(\exp\left(2 \beta\right) + 1\right) + \sqrt{\upsilon^2 \left(\exp\left(2 \beta\right) - 1\right)^2 + 4 \exp\left(2 \beta\right)}}{2 \exp\left(\beta\right) \left(1 - \upsilon\right)}\right).
\end{equation*}
Hence, the desired cdf can be written as
\begin{equation}
\begin{split}
&P\left(\psi \le 2 \log\left(\frac{\upsilon \left(\exp\left(2 \beta\right) + 1\right) + \sqrt{\upsilon^2 \left(\exp\left(2 \beta\right) - 1\right)^2 + 4 \exp\left(2 \beta\right)}}{2 \exp\left(\beta\right) \left(1 - \upsilon\right)}\right)\right) \\
=& \int_{-\infty}^{\infty} \int_{-\infty}^{2 \log\left(\frac{\upsilon \left(\exp\left(2 \beta\right) + 1\right) + \sqrt{\upsilon^2 \left(\exp\left(2 \beta\right) - 1\right)^2 + 4 \exp\left(2 \beta\right)}}{2 \exp\left(\beta\right) \left(1 - \upsilon\right)}\right)} \mathcal{N}(\psi; \mu_\psi, \sigma_\psi^2) \mathcal{N}(\beta; \mu_\beta, \sigma_\beta^2) \text{d}\psi \text{d}\beta \\
=& \int_{-\infty}^{\infty} \mathcal{N}(\beta; \mu_\beta, \sigma_\beta^2) \, \Phi\left(2 \log\left(\frac{\upsilon \left(\exp\left(2 \beta\right) + 1\right) + \sqrt{\upsilon^2 \left(\exp\left(2 \beta\right) - 1\right)^2 + 4 \exp\left(2 \beta\right)}}{2 \exp\left(\beta\right) \left(1 - \upsilon\right)}\right); \mu_\psi, \sigma_\psi^2\right) \text{d}\beta.
\end{split}
\end{equation}
The pdf of the absolute risk is obtained by taking the derivative with respect to $\upsilon$:
\begin{equation}
\begin{split}
& \frac{d}{d \upsilon} \left[\int_{-\infty}^{\infty} \mathcal{N}(\beta; \mu_\beta, \sigma_\beta^2) \, \Phi\left(2 \log\left(\frac{\upsilon \left(\exp\left(2 \beta\right) + 1\right) + \sqrt{\upsilon^2 \left(\exp\left(2 \beta\right) - 1\right)^2 + 4 \exp\left(2 \beta\right)}}{2 \exp\left(\beta\right) \left(1 - \upsilon\right)}\right); \mu_\psi, \sigma_\psi^2\right) \text{d}\beta\right] \\
&= \int_{-\infty}^{\infty} \mathcal{N}(\beta; \mu_\beta, \sigma_\beta^2) \, \mathcal{N}\left(2 \log\left(\frac{\upsilon \left(\exp\left(2 \beta\right) + 1\right) + \sqrt{\upsilon^2 \left(\exp\left(2 \beta\right) - 1\right)^2 + 4 \exp\left(2 \beta\right)}}{2 \exp\left(\beta\right) \left(1 - \upsilon\right)}\right); \mu_\psi, \sigma_\psi^2\right) \\
& \hspace{3em} \times 2 \left[\frac{\exp\left(2 \beta\right) + \frac{\upsilon \left(\exp\left(2 \beta\right) - 1\right)^2}{\sqrt{\upsilon^2 \left(\exp\left(2 \beta\right) - 1\right)^2 + 4 \exp\left(2 \beta\right)}} + 1}{\upsilon \left(\exp\left(2 \beta\right) + 1\right) + \sqrt{\upsilon^2 \left(\exp\left(2 \beta\right) - 1\right)^2 + 4 \exp\left(2 \beta\right)}} + \frac{1}{1 - \upsilon}\right] \text{d}\beta.
\end{split}
\end{equation}

\subsection{Joint distribution of $p_1$ and $p_2$}
Another distribution of interest is the implied joint distribution of the two success probabilities $p_1$ and $p_2$.
This distribution will not be used to elicit the prior on $\psi$ which is the reason why we only derive the pdf and not the cdf.
The model parameters $\beta$ and $\psi$ are related to $p_1$ and $p_2$ as follows:
\begin{align*}
\log\left(\frac{p_1}{1 - p_1}\right) &= \beta - \frac{\psi}{2} \\
\log\left(\frac{p_2}{1 - p_2}\right) &= \beta + \frac{\psi}{2}.
\end{align*}
Hence, the inverse transformation is given by:
\begin{align*}
\beta &= \frac{1}{2} \log\left(\frac{p_1}{1 - p_1}\right) + \frac{1}{2} \log\left(\frac{p_2}{1 - p_2}\right) \\
\psi &= \log\left(\frac{p_2}{1 - p_2}\right) - \log\left(\frac{p_1}{1 - p_1}\right).
\end{align*}
The corresponding Jacobian is:
\begin{align*}
\lvert J \rvert &= \left\lvert \begin{pmatrix}
\frac{\partial \beta}{\partial p_1} & \frac{\partial \beta}{\partial p_2} \\
\frac{\partial \psi}{\partial p_1} & \frac{\partial \psi}{\partial p_2}
\end{pmatrix} \right\rvert \\
&= \left\lvert \begin{pmatrix}
\frac{1}{2} \frac{1}{p_1 (1 - p_1)} & \frac{1}{2} \frac{1}{p_2 (1 - p_2)}  \\
-\frac{1}{p_1 (1 - p_1)} & \frac{1}{p_2 (1 - p_2)}
\end{pmatrix} \right\rvert \\
&= \frac{1}{p_1 p_2 (1 - p_1) (1 - p_2)}.
\end{align*}
Therefore, the joint pdf of $p_1$ and $p_2$ is given by:
\begin{equation}
\begin{split}
p(p_1, p_2) &= \frac{1}{p_1 p_2 (1 - p_1) (1 - p_2)} \, \mathcal{N}\left(\frac{1}{2} \left[\log\left(\frac{p_1}{1 - p_1}\right) + \log\left(\frac{p_2}{1 - p_2}\right)\right]; \mu_\beta, \sigma_\beta^2\right) \, \\
& \hspace{3em} \times \mathcal{N}\left(\log\left(\frac{p_2}{1 - p_2}\right) - \log\left(\frac{p_1}{1 - p_1}\right); \mu_\psi, \sigma_\psi^2\right).
\end{split}
\end{equation}

\subsection{Marginal distribution of $p_1$}
The marginal distribution of $p_1$ is given by:
\begin{equation}
\begin{split}
p(p_1) &= \int_{0}^{1} p(p_1, p_2^\prime) \text{d}p_2^\prime \\
&=  \int_{0}^{1} \frac{1}{p_1 p_2^\prime (1 - p_1) (1 - p_2^\prime)} \, \mathcal{N}\left(\frac{1}{2} \left[\log\left(\frac{p_1}{1 - p_1}\right) + \log\left(\frac{p_2^\prime}{1 - p_2^\prime}\right)\right]; \mu_\beta, \sigma_\beta^2\right) \, \\
& \hspace{3em} \times \mathcal{N}\left(\log\left(\frac{p_2^\prime}{1 - p_2^\prime}\right) - \log\left(\frac{p_1}{1 - p_1}\right); \mu_\psi, \sigma_\psi^2\right) \text{d}p_2^\prime.
\end{split}
\end{equation}

\subsection{Marginal distribution of $p_2$}
The marginal distribution of $p_2$ is given by:
\begin{equation}
\begin{split}
p(p_2) &= \int_{0}^{1} p(p_1^\prime, p_2) \text{d}p_1^\prime \\
&=  \int_{0}^{1} \frac{1}{p_1^\prime p_2 (1 - p_1^\prime) (1 - p_2)} \, \mathcal{N}\left(\frac{1}{2} \left[\log\left(\frac{p_1^\prime}{1 - p_1^\prime}\right) + \log\left(\frac{p_2}{1 - p_2}\right)\right]; \mu_\beta, \sigma_\beta^2\right) \, \\
& \hspace{3em} \times \mathcal{N}\left(\log\left(\frac{p_2}{1 - p_2}\right) - \log\left(\frac{p_1^\prime}{1 - p_1^\prime}\right); \mu_\psi, \sigma_\psi^2\right) \text{d}p_1^\prime.
\end{split}
\end{equation}

\subsection{Conditional distribution of $p_2$ given $p_1$}
Another distribution of interest is the conditional distribution of the second success probability $p_2$ given a particular value of $p_1$.  This distribution will not be used for prior elicitation which is the reason why we only present the expression for the pdf which is given by:
\begin{equation}
\begin{split}
p&(p_2 \mid p_1) = \frac{p(p_1, p_2)}{\int_{0}^{1} p(p_1, p_2^\prime) \text{d}p_2^\prime} \\
&= \frac{\frac{1}{p_2 (1 - p_2)} \, \mathcal{N}\left(\frac{1}{2} \left[\log\left(\frac{p_1}{1 - p_1}\right) + \log\left(\frac{p_2}{1 - p_2}\right)\right]; \mu_\beta, \sigma_\beta^2\right) \, \mathcal{N}\left(\log\left(\frac{p_2}{1 - p_2}\right) - \log\left(\frac{p_1}{1 - p_1}\right); \mu_\psi, \sigma_\psi^2\right)}{\int_{0}^{1} \frac{1}{ p_2^\prime (1 - p_2^\prime)} \, \mathcal{N}\left(\frac{1}{2} \left[\log\left(\frac{p_1}{1 - p_1}\right) + \log\left(\frac{p_2^\prime}{1 - p_2^\prime}\right)\right]; \mu_\beta, \sigma_\beta^2\right) \,  \mathcal{N}\left(\log\left(\frac{p_2^\prime}{1 - p_2^\prime}\right) - \log\left(\frac{p_1}{1 - p_1}\right); \mu_\psi, \sigma_\psi^2\right) \text{d}p_2^\prime}.
\end{split}
\end{equation}

\subsection{Implied distributions for truncated priors on the log odds ratio}
Note that the above expressions can be all easily modified in case the prior on the log odds ratio $\psi$ is a truncated normal distribution (e.g., restricting $\psi$ to be larger/smaller than zero) which is the case for the hypotheses $\mathcal{H}_+$ and $\mathcal{H}_-$. In this case, the normal prior density function and cumulative distribution function for $\psi$ simply need to be replaced by the truncated versions. For the implied log-normal prior on the odds ratio, the truncation bounds simply need to be exponentiated to obtain the truncation bounds with respect to the log-normal prior.

\section{Laplace approximation details}
The Laplace approximations require first-order and second-order derivatives. Let us first state explicitly the functions for which we need to find the derivatives.
For $\mathcal{H}_0$ we have:
\begin{equation}
\begin{split}
l_0^\ast(\beta) &= \log\left\{p(y \mid \beta) \, \pi_0(\beta)\right\} \\
&= (y_1 + y_2) \log\left(\frac{\exp(\beta)}{1 + \exp(\beta)}\right) + (n_1 + n_2 - y_1 - y_2) \log\left(1 - \frac{\exp(\beta)}{1 + \exp(\beta)}\right) \\
& \hspace{3em} - \frac{1}{2}\log\left(2 \pi \sigma^2_\beta\right) - \frac{1}{2 \sigma^2_\beta} (\beta - \mu_\beta)^2.
\end{split}
\end{equation}
For $\mathcal{H}_1$ we have:
\begin{equation}
\begin{split}
l^\ast(\beta, \psi) &= \log\left\{p(y \mid \beta, \psi) \, \pi(\beta, \psi)\right\} \\
&= y_1 \log\left(\frac{\exp(\beta - \frac{\psi}{2})}{1 + \exp(\beta - \frac{\psi}{2})}\right) + (n_1 - y_1) \log\left(1 - \frac{\exp(\beta - \frac{\psi}{2})}{1 + \exp(\beta - \frac{\psi}{2})}\right) \\
& \hspace{3em} + y_2 \log\left(\frac{\exp(\beta + \frac{\psi}{2})}{1 + \exp(\beta + \frac{\psi}{2})}\right) + (n_2 - y_2) \log\left(1 - \frac{\exp(\beta + \frac{\psi}{2})}{1 + \exp(\beta + \frac{\psi}{2})}\right) \\
& \hspace{3em} - \frac{1}{2}\log\left(2 \pi \sigma^2_\beta\right) - \frac{1}{2 \sigma^2_\beta} (\beta - \mu_\beta)^2 - \frac{1}{2}\log\left(2 \pi \sigma^2_\psi\right) - \frac{1}{2 \sigma^2_\psi} (\psi - \mu_\psi)^2.
\end{split}
\end{equation}
For $\mathcal{H}_+$ we have:
\begin{equation}
\begin{split}
l_+^\ast(\beta, \xi) &= \log\left\{p(y \mid \beta, \xi) \, \pi_+(\beta, \xi)\right\} \\
&=y_1 \log\left(\frac{\exp(\beta - \frac{\exp(\xi)}{2})}{1 + \exp(\beta - \frac{\exp(\xi)}{2})}\right) + (n_1 - y_1) \log\left(1 - \frac{\exp(\beta - \frac{\exp(\xi)}{2})}{1 + \exp(\beta - \frac{\exp(\xi)}{2})}\right) \\
& \hspace{3em} + y_2 \log\left(\frac{\exp(\beta + \frac{\exp(\xi)}{2})}{1 + \exp(\beta + \frac{\exp(\xi)}{2})}\right) + (n_2 - y_2) \log\left(1 - \frac{\exp(\beta + \frac{\exp(\xi)}{2})}{1 + \exp(\beta + \frac{\exp(\xi)}{2})}\right) \\
& \hspace{3em} - \frac{1}{2}\log\left(2 \pi \sigma^2_\beta\right) - \frac{1}{2 \sigma^2_\beta} (\beta - \mu_\beta)^2 \\
& \hspace{3em} - \frac{1}{2}\log\left(2 \pi \sigma^2_\psi\right) - \frac{1}{2 \sigma^2_\psi} (\exp(\xi) - \mu_\psi)^2 - \log(1 - \Phi\left(0; \mu_\psi, \sigma^2_\psi\right)) + \xi.
\end{split}
\end{equation}
Finally, for $\mathcal{H}_-$ we have
\begin{equation}
\begin{split}
l_-^\ast(\beta, \xi) &= \log\left\{p(y \mid \beta, \xi) \, \pi_-(\beta, \xi)\right\} \\
&= y_1 \log\left(\frac{\exp(\beta + \frac{\exp(\xi)}{2})}{1 + \exp(\beta + \frac{\exp(\xi)}{2})}\right) + (n_1 - y_1) \log\left(1 - \frac{\exp(\beta + \frac{\exp(\xi)}{2})}{1 + \exp(\beta + \frac{\exp(\xi)}{2})}\right) \\
& \hspace{3em} + y_2 \log\left(\frac{\exp(\beta - \frac{\exp(\xi)}{2})}{1 + \exp(\beta - \frac{\exp(\xi)}{2})}\right) + (n_2 - y_2) \log\left(1 - \frac{\exp(\beta - \frac{\exp(\xi)}{2})}{1 + \exp(\beta - \frac{\exp(\xi)}{2})}\right) \\
& \hspace{3em} - \frac{1}{2}\log\left(2 \pi \sigma^2_\beta\right) - \frac{1}{2 \sigma^2_\beta} (\beta - \mu_\beta)^2 \\
& \hspace{3em} - \frac{1}{2}\log\left(2 \pi \sigma^2_\psi\right) - \frac{1}{2 \sigma^2_\psi} (-\exp(\xi) - \mu_\psi)^2 - \log(\Phi\left(0; \mu_\psi, \sigma^2_\psi\right)) + \xi.
\end{split}
\end{equation}

\subsection{First-order derivatives}
The first-order derivatives are used to find the modes for the Laplace approximations.
As shown below, we can find these derivatives analytically; however, setting the derivatives equal to zero and solving for the parameters is not straightforward. Nevertheless, having these derivatives is useful not only as an intermediate step to finding the second-order derivatives but also for finding the modes: This allows us to provide numerical optimizers with the analytic expressions for the derivatives which can increase speed and accuracy for numerically finding the modes of the relevant functions.

The first-order derivative for $l_0(\beta)$ is given by:
\begin{equation}
\frac{d}{d\beta} \, l_0^\ast(\beta) = \frac{y_1 + y_2 - (n_1 + n_2 - y_1 - y_2) \exp(\beta)}{1 + \exp(\beta)} - \frac{\beta - \mu_\beta}{\sigma^2_\beta}.
\end{equation}
The first-order partial derivatives for $l^\ast(\beta, \psi)$ are given by
\begin{equation}
\frac{\partial}{\partial \beta} \, l^\ast(\beta, \psi) = \, \frac{y_1 - (n_1 - y_1) \exp(\beta - \frac{\psi}{2}) }{1 + \exp(\beta - \frac{\psi}{2})} + \frac{y_2 - (n_2 - y_2) \exp(\beta + \frac{\psi}{2}) }{1 + \exp(\beta + \frac{\psi}{2})} - \frac{\beta - \mu_\beta}{\sigma^2_\beta},
\end{equation}
and
\begin{equation}
\frac{\partial}{\partial \psi} \, l^\ast(\beta, \psi) =  \frac{1}{2} \left(\frac{(n_1 - y_1) \exp(\beta - \frac{\psi}{2}) - y_1}{1 + \exp(\beta - \frac{\psi}{2})} + \frac{y_2 -  (n_2 - y_2) \exp(\beta + \frac{\psi}{2})}{1 + \exp(\beta + \frac{\psi}{2})}\right) - \frac{\psi - \mu_\psi}{\sigma^2_\psi}.
\end{equation}
The first-order partial derivatives for $l_+^\ast(\beta, \xi)$ are given by:
\begin{equation}
\frac{\partial}{\partial \beta} \, l_+^\ast(\beta, \xi) =  \frac{y_1 - (n_1 - y_1) \exp\left(\beta - \frac{\exp(\xi)}{2}\right) }{1 + \exp\left(\beta - \frac{\exp(\xi)}{2}\right)} + \frac{y_2 - (n_2 - y_2) \exp\left(\beta + \frac{\exp(\xi)}{2}\right) }{1 + \exp\left(\beta + \frac{\exp(\xi)}{2}\right)} - \frac{\beta - \mu_\beta}{\sigma^2_\beta},
\end{equation}
and
\begin{equation}
\begin{split}
\frac{\partial}{\partial \xi} \, l_+^\ast(\beta, \xi) &=  \frac{\exp(\xi)}{2} \left(\frac{(n_1 - y_1) \exp(\beta - \frac{\exp(\xi)}{2}) - y_1}{1 + \exp(\beta - \frac{\exp(\xi)}{2})} + \frac{y_2 -  (n_2 - y_2) \exp(\beta + \frac{\exp(\xi)}{2})}{1 + \exp(\beta + \frac{\exp(\xi)}{2})}\right) \\
& \hspace{3em} - \exp(\xi) \frac{\exp(\xi) - \mu_\psi}{\sigma^2_\psi} + 1.
\end{split}
\end{equation}
The first-order partial derivatives for $l_-^\ast(\beta, \xi)$ are given by:
\begin{equation}
\frac{\partial}{\partial \beta} \, l_-^\ast(\beta, \xi) = \, \frac{y_1 - (n_1 - y_1) \exp\left(\beta + \frac{\exp(\xi)}{2}\right) }{1 + \exp\left(\beta + \frac{\exp(\xi)}{2}\right)} + \frac{y_2 - (n_2 - y_2) \exp\left(\beta - \frac{\exp(\xi)}{2}\right) }{1 + \exp\left(\beta - \frac{\exp(\xi)}{2}\right)} - \frac{\beta - \mu_\beta}{\sigma^2_\beta},
\end{equation}
and
\begin{equation}
\begin{split}
\frac{\partial}{\partial \xi} \, l_-^\ast(\beta, \xi) &=  \frac{\exp(\xi)}{2} \left(\frac{y_1 - (n_1 - y_1) \exp(\beta + \frac{\exp(\xi)}{2})}{1 + \exp(\beta + \frac{\exp(\xi)}{2})} + \frac{(n_2 - y_2) \exp(\beta - \frac{\exp(\xi)}{2}) - y_2}{1 + \exp(\beta - \frac{\exp(\xi)}{2})}\right) \\
& \hspace{3em} + \exp(\xi) \frac{-\exp(\xi) - \mu_\psi}{\sigma^2_\psi} + 1.
\end{split}
\end{equation}

\subsection{Second-order derivatives}
For the Laplace approximations, we also need the inverse of the negative Hessians. The Hessian is the matrix with the second-order partial derivatives which is the reason why we now present expressions for the second-order partial derivatives.
Note that under all hypotheses there are either one or two parameters. Hence, the Hessians will be at most 2 by 2 matrices. For matrices up to 2 by 2, it is straightforward to find the inverse and the determinant which makes it easy to obtain the quantities needed for the Laplace approximations once we have the required derivatives.

For $l_0^\ast(\beta)$, there is only one parameter and the second-order derivative is given by:
\begin{equation}
\frac{d^2}{d\beta^2} \, l_0^\ast(\beta) = - \frac{(n_1 + n_2) \exp(\beta)}{\left(1 + \exp(\beta)\right)^2} - \frac{1}{\sigma^2_\beta}.
\end{equation}
For $l^\ast(\beta, \psi)$ the second-order partial derivatives are given by
\begin{equation}
\frac{\partial^2}{\partial \beta^2} \, l^\ast(\beta, \psi) = - \frac{n_1 \exp(\beta - \frac{\psi}{2})}{\left(1 + \exp(\beta - \frac{\psi}{2})\right)^2} - \frac{n_2 \exp(\beta + \frac{\psi}{2})}{\left(1 + \exp(\beta + \frac{\psi}{2})\right)^2}  - \frac{1}{\sigma^2_\beta}, 
\end{equation}
and
\begin{equation}
\frac{\partial^2}{\partial \beta \partial \psi} \, l^\ast(\beta, \psi) = \frac{1}{2} \left(\frac{n_1 \exp(\beta - \frac{\psi}{2})}{\left(1 + \exp(\beta - \frac{\psi}{2})\right)^2} - \frac{n_2 \exp(\beta + \frac{\psi}{2})}{\left(1 + \exp(\beta + \frac{\psi}{2})\right)^2}\right),
\end{equation}
and
\begin{equation}
\frac{\partial^2}{\partial \psi^2} \, l^\ast(\beta, \psi) =  - \frac{1}{4} \left(\frac{n_1 \exp(\beta - \frac{\psi}{2})}{\left(1 + \exp(\beta - \frac{\psi}{2})\right)^2} + \frac{n_2 \exp(\beta + \frac{\psi}{2})}{\left(1 + \exp(\beta + \frac{\psi}{2})\right)^2}\right) - \frac{1}{\sigma^2_\psi}.
\end{equation}
For $l_+^\ast(\beta, \xi)$ the second-order partial derivatives are given by
\begin{equation}
\frac{\partial^2}{\partial \beta^2} \, l_+^\ast(\beta, \xi) =  - \frac{n_1 \exp\left(\beta - \frac{\exp(\xi)}{2}\right)}{\left(1 + \exp\left(\beta - \frac{\exp(\xi)}{2}\right)\right)^2} - \frac{n_2 \exp\left(\beta + \frac{\exp(\xi)}{2}\right)}{\left(1 + \exp\left(\beta + \frac{\exp(\xi)}{2}\right)\right)^2}  - \frac{1}{\sigma^2_\beta}, 
\end{equation}
and
\begin{equation}
\frac{\partial^2}{\partial \beta \partial \xi} \, l_+^\ast(\beta, \xi) =  \frac{\exp(\xi)}{2} \left(\frac{n_1 \exp\left(\beta - \frac{\exp(\xi)}{2}\right)}{\left(1 + \exp\left(\beta - \frac{\exp(\xi)}{2}\right)\right)^2} - \frac{n_2 \exp\left(\beta + \frac{\exp(\xi)}{2}\right)}{\left(1 + \exp\left(\beta + \frac{\exp(\xi)}{2}\right)\right)^2}\right),
\end{equation}
and
\begin{equation}
\begin{split}
\frac{\partial^2}{\partial \xi^2} \, l_+^\ast(\beta, \xi) &=  \frac{\exp(\xi)}{2} \Bigg(\frac{(n_1 - y_1) \exp(\beta - \frac{\exp(\xi)}{2}) - y_1}{1 + \exp(\beta - \frac{\exp(\xi)}{2})} + \frac{y_2 -  (n_2 - y_2) \exp(\beta + \frac{\exp(\xi)}{2})}{1 + \exp(\beta + \frac{\exp(\xi)}{2})} \\
& \hspace{6em} - \frac{1}{2} \exp(\xi) \frac{n_1 \exp\left(\beta - \frac{\exp(\xi)}{2}\right)}{\left(1 + \exp\left(\beta - \frac{\exp(\xi)}{2}\right)\right)^2} - \frac{1}{2} \exp(\xi) \frac{n_2 \exp\left(\beta + \frac{\exp(\xi)}{2}\right)}{\left(1 + \exp\left(\beta + \frac{\exp(\xi)}{2}\right)\right)^2}\Bigg) \\
& \hspace{3em} - \exp(\xi) \frac{2 \exp(\xi) - \mu_\psi}{\sigma^2_\psi}.
\end{split}
\end{equation}
For $l_-^\ast(\beta, \xi)$ the second-order partial derivatives are given by
\begin{equation}
\frac{\partial^2}{\partial \beta^2} \, l_-^\ast(\beta, \xi) =  - \frac{n_1 \exp\left(\beta + \frac{\exp(\xi)}{2}\right)}{\left(1 + \exp\left(\beta + \frac{\exp(\xi)}{2}\right)\right)^2} - \frac{n_2 \exp\left(\beta - \frac{\exp(\xi)}{2}\right)}{\left(1 + \exp\left(\beta - \frac{\exp(\xi)}{2}\right)\right)^2}  - \frac{1}{\sigma^2_\beta}, 
\end{equation}
and
\begin{equation}
\frac{\partial^2}{\partial \beta \partial \xi} \, l_-^\ast(\beta, \xi) =  - \frac{\exp(\xi)}{2} \left(\frac{n_1 \exp\left(\beta + \frac{\exp(\xi)}{2}\right)}{\left(1 + \exp\left(\beta + \frac{\exp(\xi)}{2}\right)\right)^2} - \frac{n_2 \exp\left(\beta - \frac{\exp(\xi)}{2}\right)}{\left(1 + \exp\left(\beta - \frac{\exp(\xi)}{2}\right)\right)^2}\right),
\end{equation}
and
\begin{equation}
\begin{split}
\frac{\partial^2}{\partial \xi^2} \, l_-^\ast(\beta, \xi) &= - \frac{\exp(\xi)}{2} \Bigg(\frac{(n_1 - y_1) \exp(\beta + \frac{\exp(\xi)}{2}) - y_1}{1 + \exp(\beta + \frac{\exp(\xi)}{2})} + \frac{y_2 -  (n_2 - y_2) \exp(\beta - \frac{\exp(\xi)}{2})}{1 + \exp(\beta - \frac{\exp(\xi)}{2})} \\
& \hspace{6em} - \frac{1}{2} \exp(\xi) \frac{n_1 \exp\left(\beta + \frac{\exp(\xi)}{2}\right)}{\left(1 + \exp\left(\beta + \frac{\exp(\xi)}{2}\right)\right)^2} - \frac{1}{2} \exp(\xi) \frac{n_2 \exp\left(\beta - \frac{\exp(\xi)}{2}\right)}{\left(1 + \exp\left(\beta - \frac{\exp(\xi)}{2}\right)\right)^2}\Bigg) \\
& \hspace{3em} + \exp(\xi) \frac{2 \exp(\xi) - \mu_\psi}{\sigma^2_\psi}.
\end{split}
\end{equation}

\subsection{Hessians}
Having derived the relevant second-order partial derivatives, we can simply build the Hessian matrices of interest by inserting the relevant expressions.
Next, we present symbolically the Hessians of interest, that is, we show which of the second-order partial derivatives need to be inserted where. Note that we omit the one for $\mathcal{H}_0$ since this is a single number which is simply the second-order derivative of $l_0^\ast(\beta)$.

The Hessian for $\mathcal{H}_1$ is given by:
\begin{equation}
\boldsymbol{H}_1 = 
\begin{pmatrix}
\frac{\partial^2}{\partial \beta^2} \, l^\ast(\beta, \psi) & \frac{\partial^2}{\partial \beta \partial \psi} \, l^\ast(\beta, \psi) \\
\frac{\partial^2}{\partial \beta \partial \psi} \, l^\ast(\beta, \psi) & \frac{\partial^2}{\partial \psi^2} \, l^\ast(\beta, \psi)
\end{pmatrix}.
\end{equation}
The Hessian for $\mathcal{H}_+$ is given by:
\begin{equation}
\boldsymbol{H}_+ = 
\begin{pmatrix}
\frac{\partial^2}{\partial \beta^2} \, l_+^\ast(\beta, \xi) & \frac{\partial^2}{\partial \beta \partial \xi} \, l_+^\ast(\beta, \xi) \\
\frac{\partial^2}{\partial \beta \partial \xi} \, l_+^\ast(\beta, \xi) & \frac{\partial^2}{\partial \xi^2} \, l_+^\ast(\beta, \xi)
\end{pmatrix}.
\end{equation}
The Hessian for $\mathcal{H}_-$ is given by:
\begin{equation}
\boldsymbol{H}_- = 
\begin{pmatrix}
\frac{\partial^2}{\partial \beta^2} \, l_-^\ast(\beta, \xi) & \frac{\partial^2}{\partial \beta \partial \xi} \, l_-^\ast(\beta, \xi) \\
\frac{\partial^2}{\partial \beta \partial \xi} \, l_-^\ast(\beta, \xi) & \frac{\partial^2}{\partial \xi^2} \, l_-^\ast(\beta, \xi)
\end{pmatrix}.
\end{equation}

\subsubsection{Computing the inverse of the negative Hessians}
Note that computing the inverses of the 2 by 2 negative Hessians is straightforward:
We simply need to attach minus signs to each element of the Hessians and then make use of the fact that the inverse of a 2 by 2 matrix $\boldsymbol{A} = \begin{pmatrix} a & b \\ c & d \end{pmatrix}$ is given by $\boldsymbol{A}^{-1} = \frac{1}{\det\left(\boldsymbol{A}\right)}\begin{pmatrix} d & -b \\ -c & a \end{pmatrix}$, where $\det\left(\boldsymbol{A}\right) = ad - bc$.

\section{Example 1: effectiveness of resilience training (default analysis)}
Here we present the results for the resilience training example obtained using the default prior setting.

\subsection{Prior specification}
\begin{figure}
	\centering
	\begin{tabular}{c}
		\includegraphics[width = 0.53 \textwidth]{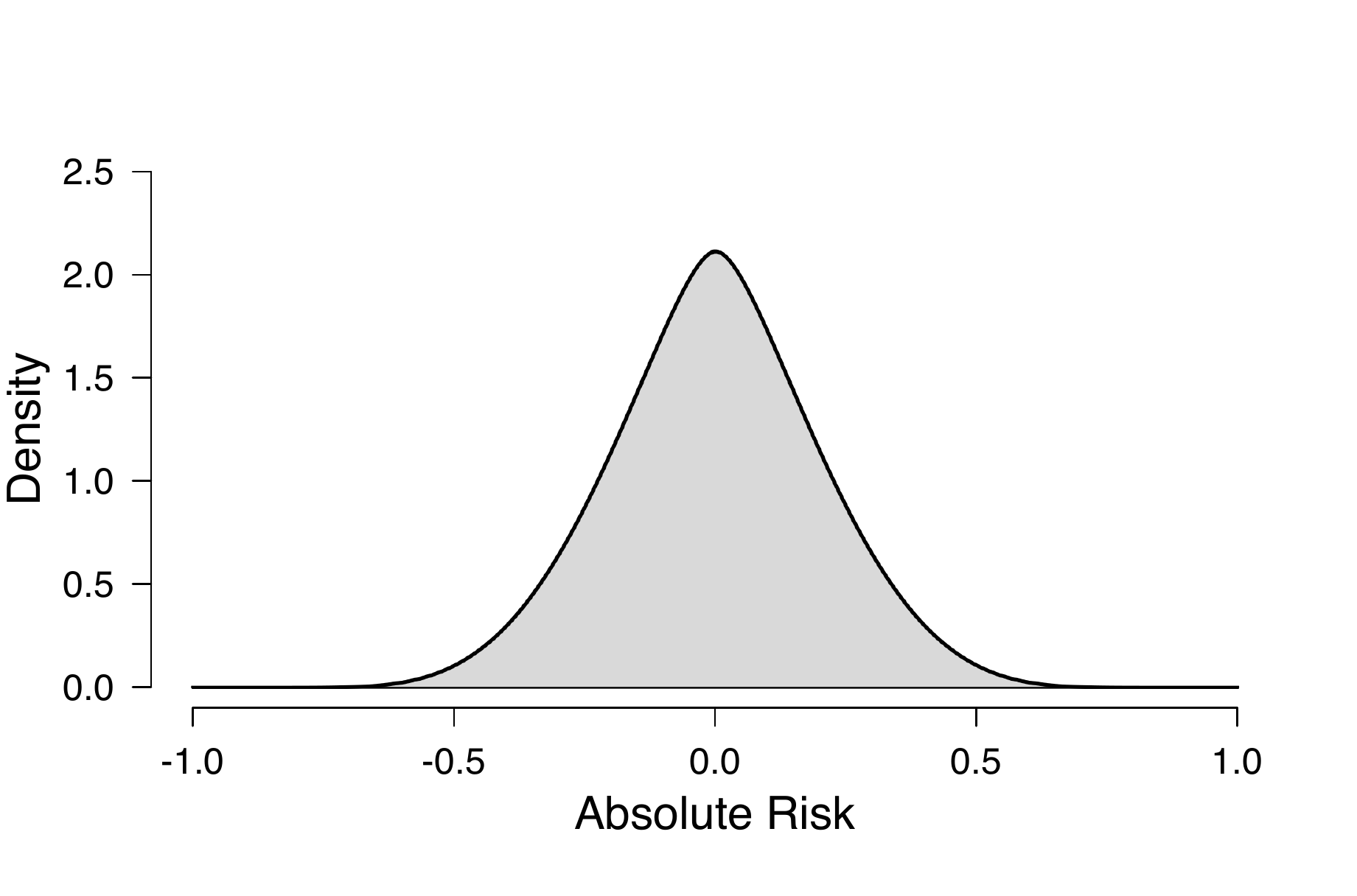} \\
		\includegraphics[width = 0.53 \textwidth]{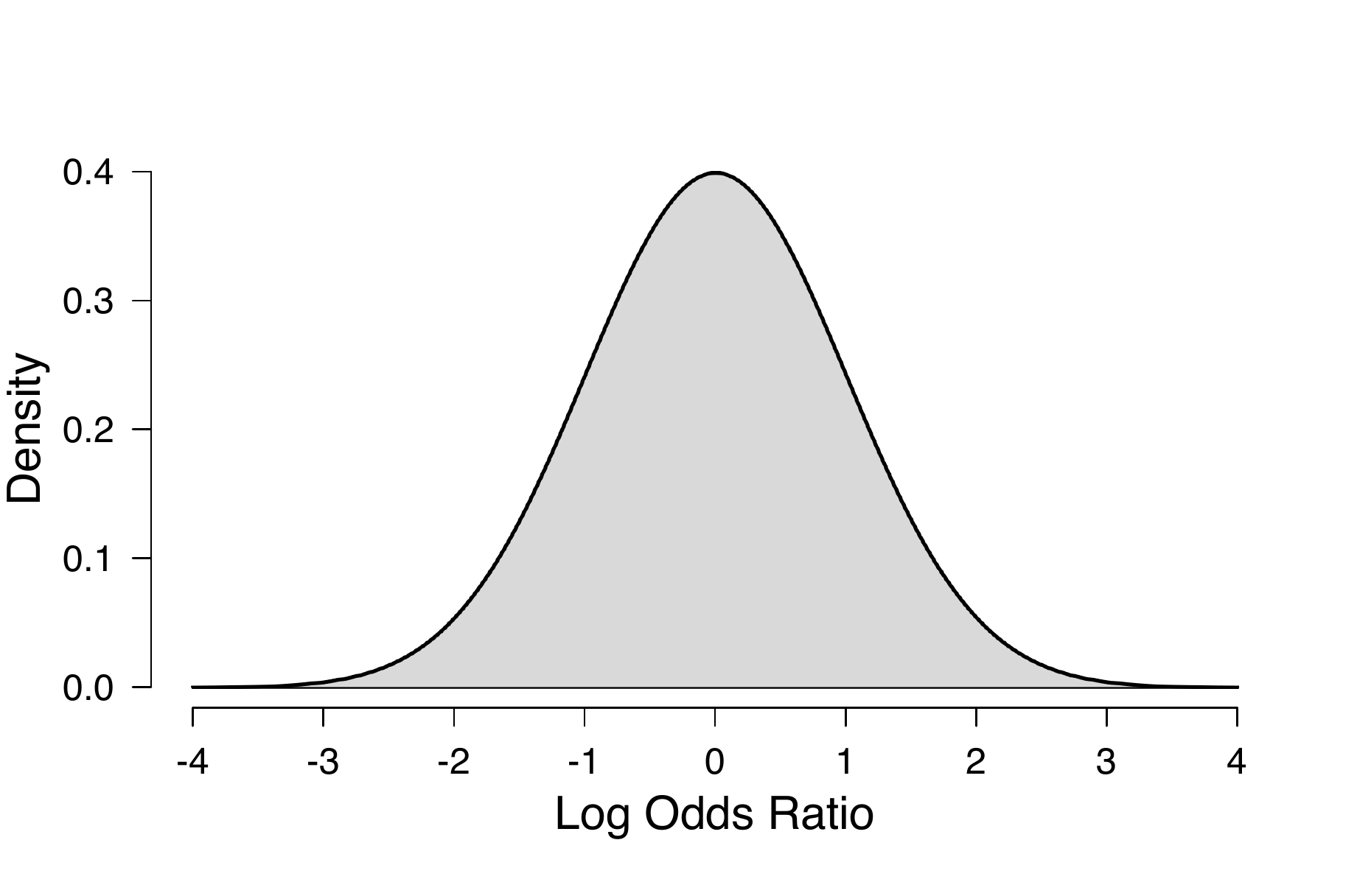} \\
		\includegraphics[width = 0.53 \textwidth]{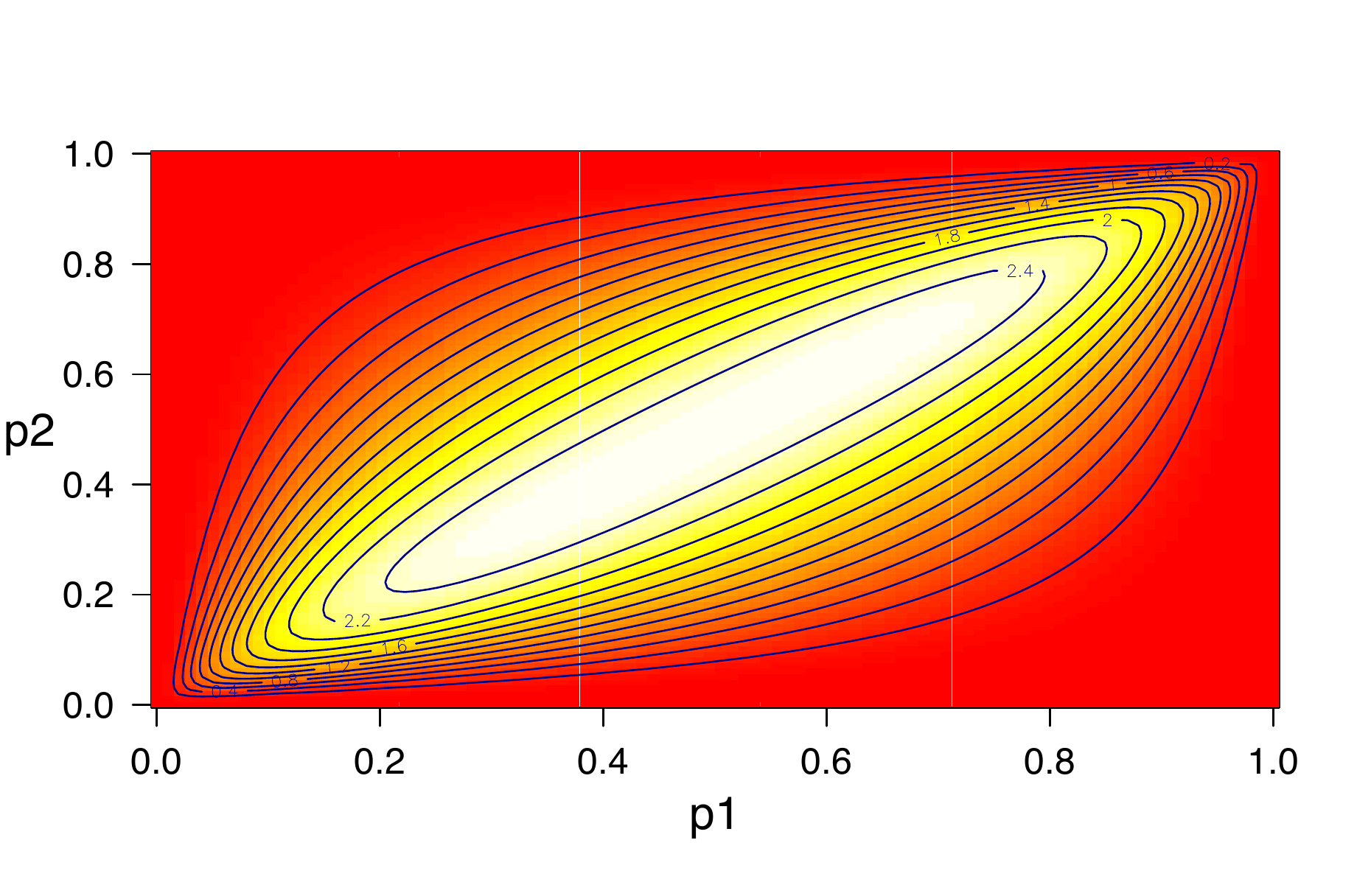}
	\end{tabular}
	\caption{Default (implied) prior distributions. The top panel displays the prior distribution for the absolute risk which corresponds to the difference between the probability of still being on the job for the trained and the non-trained employees (i.e., $p_2 - p_1$). The middle panel shows the prior distribution for the log odds ratio parameter $\psi$. The bottom panel displays the implied joint prior distribution for the success probabilities $p_1$ and $p_2$. The bottom panel illustrates that the two success probabilities are assigned dependent priors.}
	\label{fig:priors_default}
\end{figure}

We use the default prior setting in the \pkg{abtest} package that assigns both $\beta$ and $\psi$ standard normal prior distributions.
The implied prior on the absolute risk can be visualized as follows:
\begin{verbatim}
R> library("abtest")
R> plot_prior(what = "arisk")
\end{verbatim}
The resulting graph is shown in the top panel of Figure~\ref{fig:priors_default}. 
The user can also visualize the (implied) prior for other quantities. For instance, the prior on the log odds ratio (middle panel of Figure~\ref{fig:priors_default}) is obtained as follows:
\begin{verbatim}
R> plot_prior(what = "logor")
\end{verbatim}
The implied prior on the success probabilities $p_1$ and $p_2$ (bottom panel of Figure~\ref{fig:priors_default}) is obtained as follows:
\begin{verbatim}
R> plot_prior(what = "p1p2")
\end{verbatim}
The bottom panel of Figure~\ref{fig:priors_default} illustrates that there is a dependency between $p_1$ and $p_2$ which is arguably desirable \citep{Howard1998}: When one of the success probabilities is very (small) large, it is likely that the other one will also be (small) large.

\subsection{Hypothesis testing}
The \texttt{ab\_test} function can be used to conduct a Bayesian A/B test using the default prior setting as follows:
\begin{verbatim}
R> data("seqdata")
R> set.seed(1)
R> ab_default <- ab_test(data = seqdata)
\end{verbatim}
This yields the following output:
\begin{verbatim}
R> print(ab_default)

Bayesian A/B Test Results:

Bayes Factors:

BF10: 0.2767214
BF+0: 0.4890489
BF-0: 0.05778357

Prior Probabilities Hypotheses:

H+: 0.25
H-: 0.25
H0: 0.5

Posterior Probabilities Hypotheses:

H+: 0.192
H-: 0.0227
H0: 0.7853
\end{verbatim}
The first part of the output presents Bayes factors in favor of the hypotheses $\mathcal{H}_1$, $\mathcal{H}_+$, and $\mathcal{H}_-$, where the reference hypothesis (i.e., denominator of the Bayes factor) is $\mathcal{H}_0$. Since all three Bayes factors are smaller than 1, they all indicate evidence in favor of the null hypothesis of no effect. The next part of the output displays the prior probabilities of the hypotheses with non-zero prior probability. The final part of the output displays the posterior probabilities of the hypotheses with non-zero prior probability. The posterior probability of the null hypothesis $\mathcal{H}_0$ indicates that the data have increased the plausibility of the null hypothesis from $.50$ to $.79$. Furthermore, the data have decreased the plausibility of both $\mathcal{H}_+$ and $\mathcal{H}_-$.

\begin{figure}
	\centering
	\includegraphics[width = \textwidth]{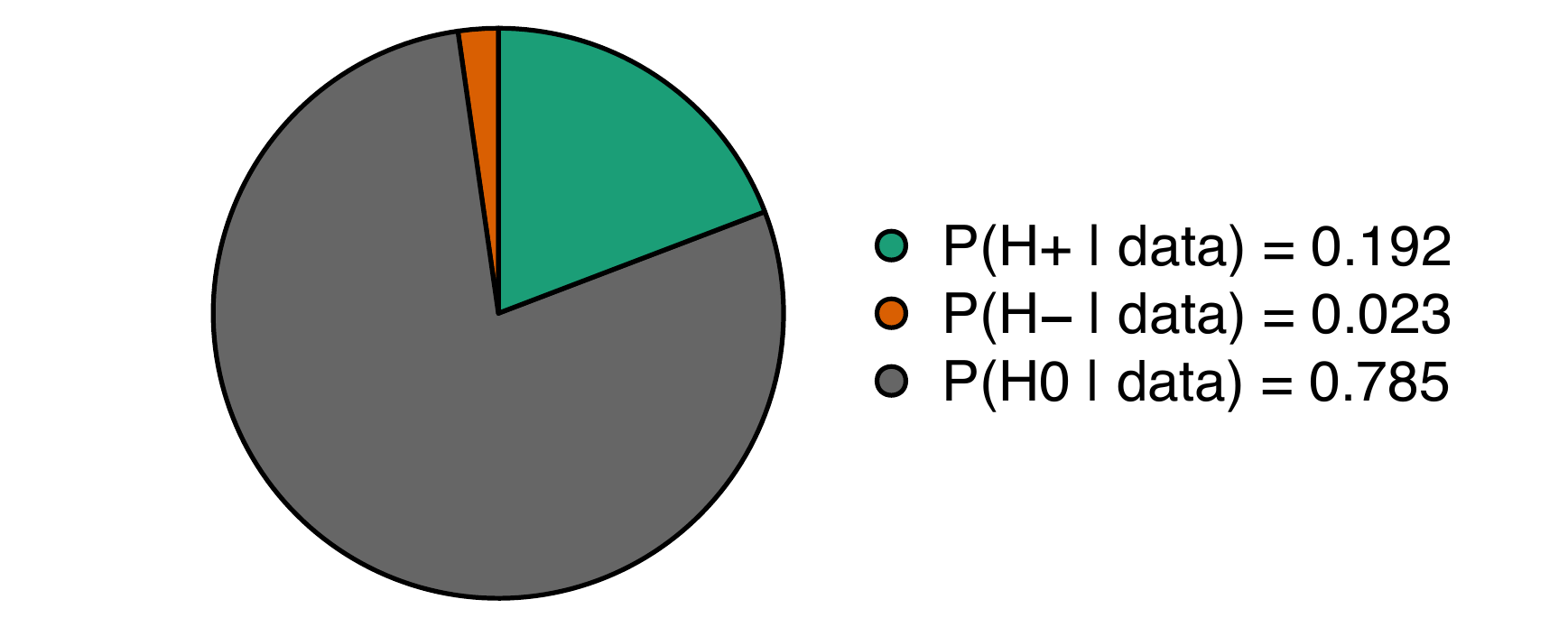}
	\caption{Posterior probabilities of the hypotheses visualized as a probability wheel.}
	\label{fig:post_probs_default}
\end{figure}
The \pkg{abtest} package allows users to visualize the posterior probabilities of the hypotheses by means of a probability wheel (Figure~\ref{fig:post_probs_default}):
\begin{verbatim}
R> prob_wheel(ab_default)
\end{verbatim}
Overall, the data support the hypothesis that the training is ineffective over the hypothesis that the training has a positive effect. The Bayes factor for $\mathcal{H}_0$ over $\mathcal{H}_+$ equals $1/0.489 \approx 2.04$; however, this indicates only anecdotal evidence \citep[Appendix I]{Jeffreys1939}.   

Since the data set is of a sequential nature, it may be of interest to consider not only the result based on all observations, but to conduct also a sequential analysis that tracks the evidential flow as a function of the total number of observations (i.e., the number of observations across both groups). This sequential analysis can be conducted as follows:
\begin{verbatim}
R> plot_sequential(ab_default, thin = 4)
\end{verbatim}
\begin{figure}
	\centering
	\includegraphics[width = \textwidth]{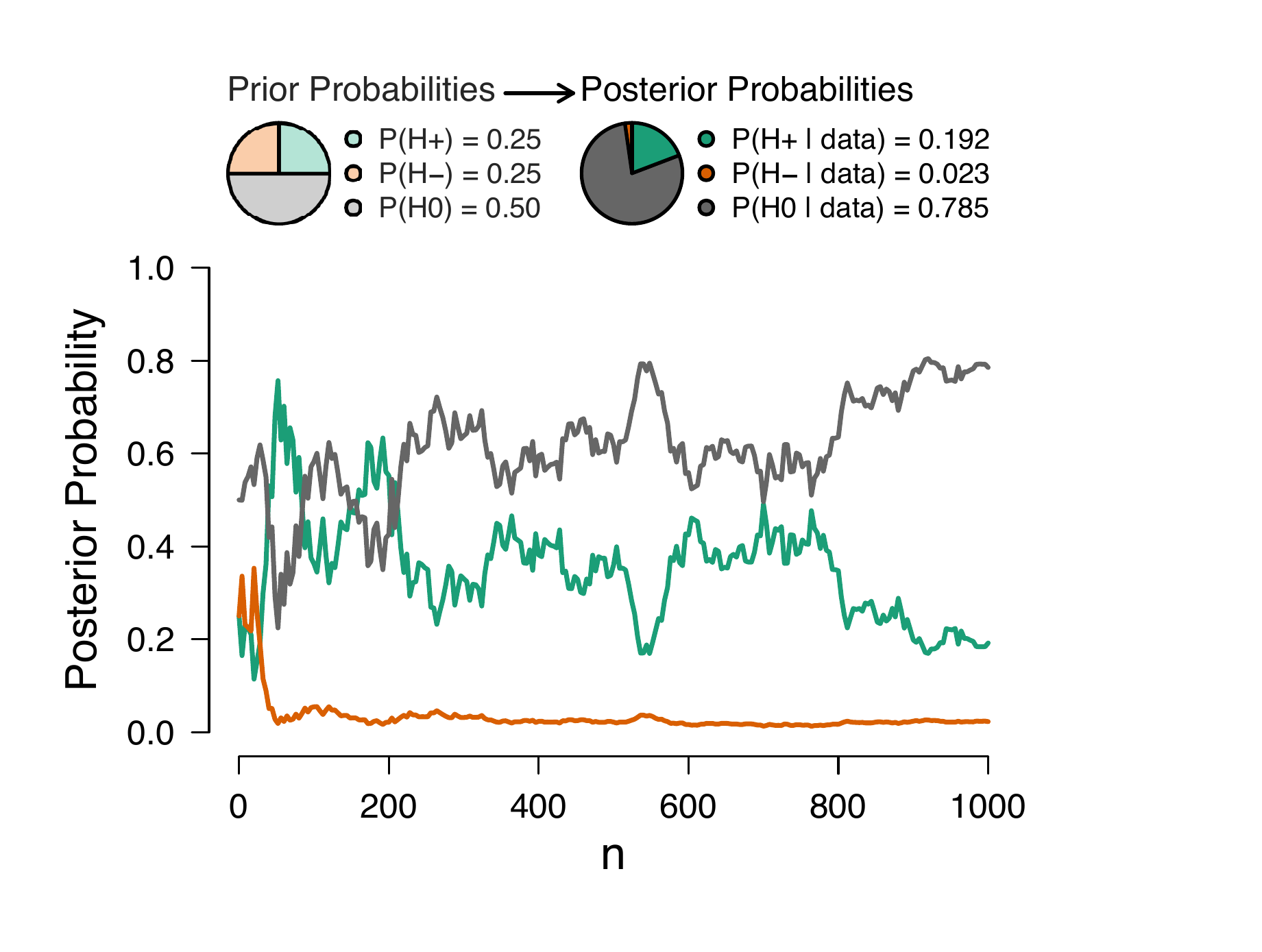}
	\caption{Sequential analysis results. The posterior probability of each hypothesis is plotted as a function of the number of observations across groups. On top, two probability wheels visualize the prior probabilities of the hypotheses and the posterior probabilities after taking into account all observations.}
	\label{fig:sequential_default}
\end{figure}
Figure~\ref{fig:sequential_default} displays the result of the sequential analysis. The sequential analysis indicates that after some initial fluctuation, adding more observations increased the probability of the null hypothesis that there is no effect of the training. 

\subsection{Parameter estimation}
The data indicate only anecdotal evidence in favor of the null hypothesis versus the hypothesis that the training is effective, leaving open the possibility that the training does have an effect. To assess this possibility one may investigate the potential size of the effect under the assumption that the effect is non-zero. For parameter estimation, we generally prefer to investigate the posterior distribution for the unconstrained alternative hypothesis $\mathcal{H}_1$.

The top panel of Figure~\ref{fig:posteriors_default} displays the posterior distribution for the absolute risk (i.e., $p_2 - p_1$) that can be obtained as follows:
\begin{verbatim}
R> plot_posterior(ab_default, what = "arisk")
\end{verbatim}
\begin{figure}
	\centering
	\begin{tabular}{c}
		\includegraphics[width = 0.56 \textwidth]{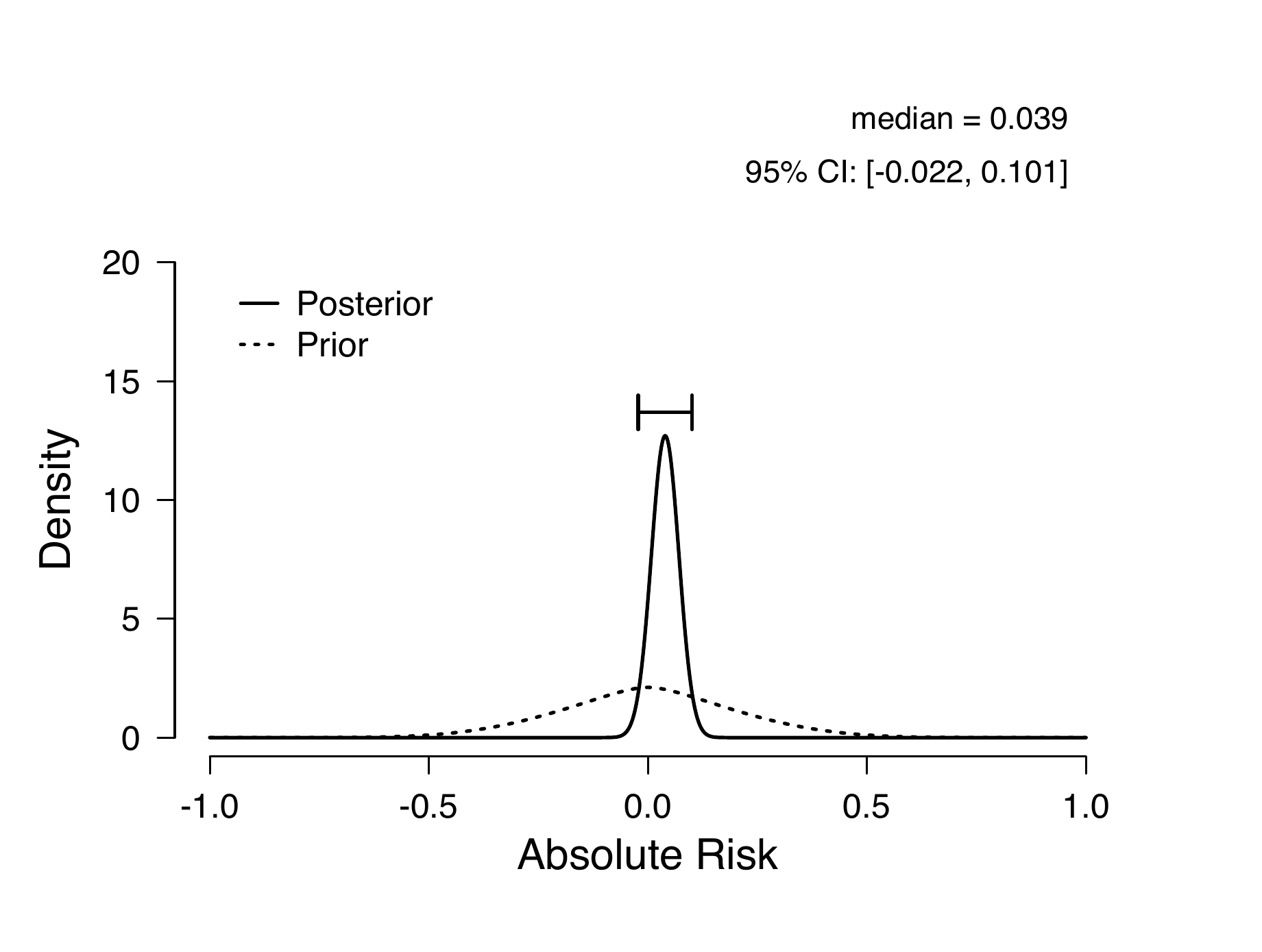} \\
		\includegraphics[width = 0.56 \textwidth]{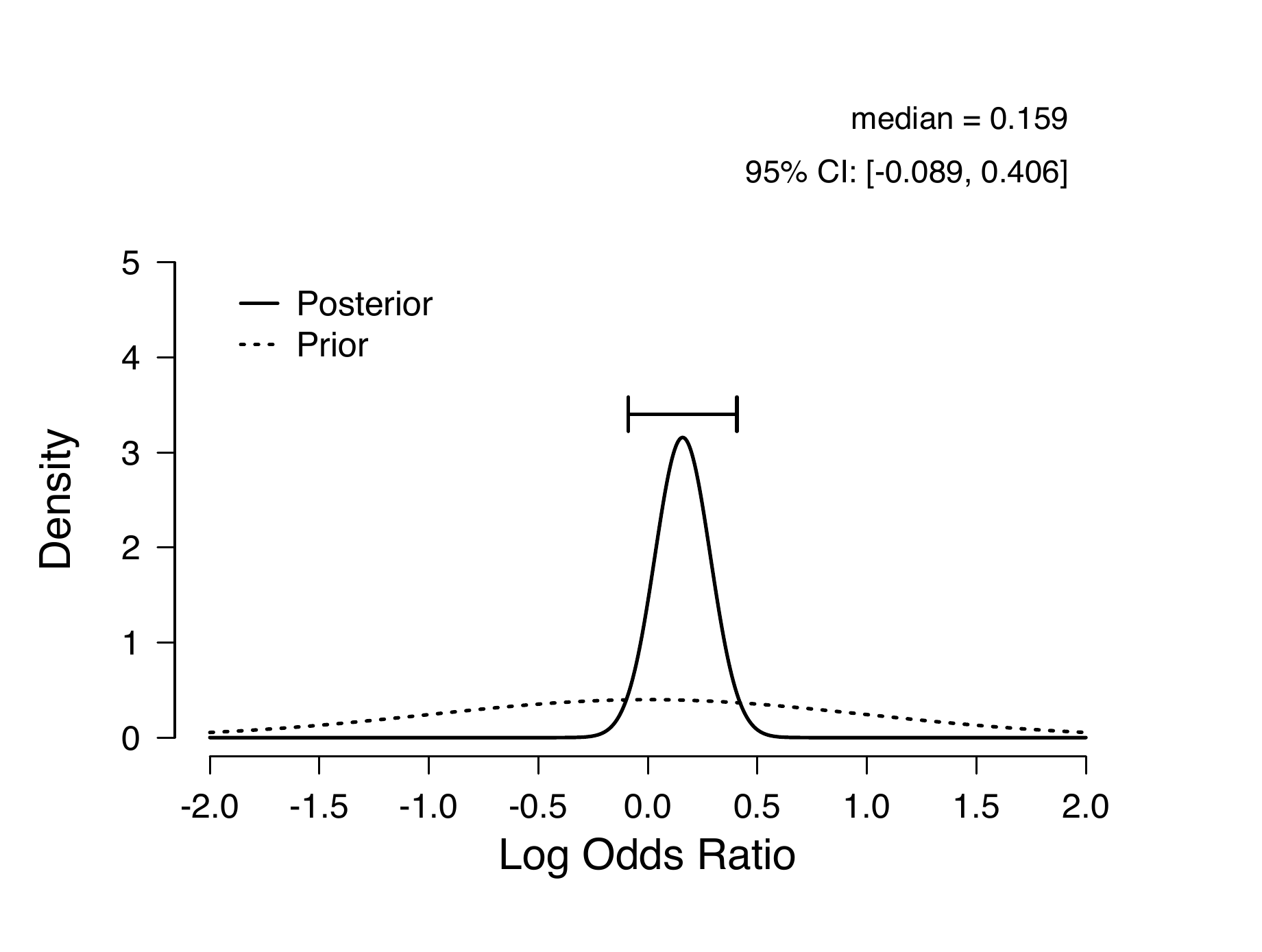} \\
		\includegraphics[width = 0.56 \textwidth]{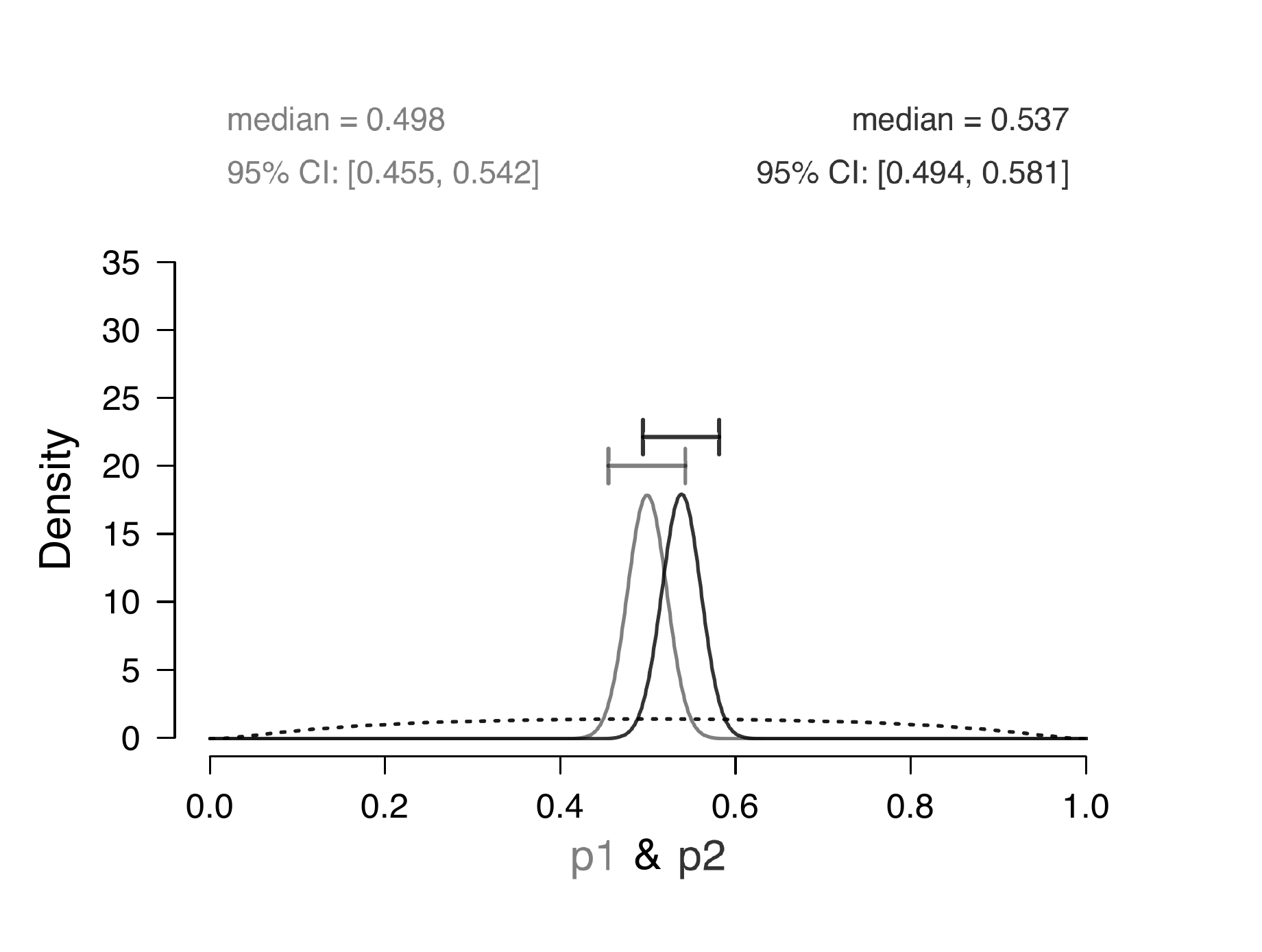}
	\end{tabular}
	\caption{(Implied) prior and posterior distributions under $\mathcal{H}_1$. The dotted lines display the prior distributions, the solid lines display the posterior distributions (with 95\% central credible intervals). The medians and the bounds of the 95\% central credible intervals are displayed on top of each panel. The top panel displays the posterior distribution for the absolute risk (i.e., $p_2 - p_1$); the middle panel shows the posterior distribution for the log odds ratio parameter $\psi$; the bottom panel displays the marginal posterior distributions for the success probabilities $p_1$ and $p_2$.}
	\label{fig:posteriors_default}
\end{figure}
The top panel of Figure~\ref{fig:posteriors_default} shows the prior distribution as a dotted line and the posterior distribution (with 95\% central credible interval) as a solid line. The plot indicates that, under the assumption that the difference between the two success probabilities is not exactly zero, the posterior median is $0.039$ and the 95\% central credible interval ranges from $-0.022$ to $0.101$.

The middle panel of Figure~\ref{fig:posteriors_default} displays the posterior distribution for the log odds ratio $\psi$ that can be obtained as follows:
\begin{verbatim}
R> plot_posterior(ab_default, what = "logor")
\end{verbatim}
The middle panel of Figure~\ref{fig:posteriors_default} indicates that, given the log odds ratio is not exactly zero, it is likely to be between $-0.089$ and $0.406$, where the posterior median is $0.159$.

It may also be of interest to consider the marginal posterior distributions of the success probabilities $p_1$ and $p_2$. This plot can be produced as follows:
\begin{verbatim}
R> plot_posterior(ab_default, what = "p1p2")
\end{verbatim}
The bottom panel of Figure~\ref{fig:posteriors_default} displays the resulting plot. In this example, $p_1$ and $p_2$ correspond to the probability of still being on the job after six month for the non-trained employees and the employees that received the training, respectively. The bottom panel of Figure~\ref{fig:posteriors_default} indicates that the posterior median for $p_1$ is $0.498$, with 95\% credible ranging from $0.455$ to $0.542$, and the posterior median for $p_2$ is $0.537$, with 95\% credible interval ranging from $0.494$ to $0.581$.

In sum, based on a default prior analysis, this fictitious data set offers anecdotal evidence in favor of the null hypothesis which states that the training is not effective over the hypothesis that the training is effective; the consultancy firm should probably continue to collect data in order to obtain more compelling evidence before deciding whether or not the training should be implemented. If the true effect is as small as 4\%, continued testing will ultimately show compelling evidence for $\mathcal{H}_+$ over $\mathcal{H}_0$. Note that continued testing is trivial in the Bayesian framework: the results can simply be updated as new observations arrive. 

\end{document}